\documentclass[twocolumn,superscriptaddress ]{revtex4-1}

\usepackage{graphicx}
\usepackage{hyperref}
\usepackage{amsmath,amssymb}
\usepackage{epstopdf}
\usepackage{subcaption}
\usepackage{color}
\graphicspath{{figs/}}
\begin{document}
\title{Scaling properties of mono-layer graphene away from the Dirac point}

\author{M. N. Najafi*}
\affiliation{Department of Physics, University of Mohaghegh Ardabili, P.O. Box 179, Ardabil, Iran}
\email{morteza.nattagh@gmail.com}

\author{N. Ahadpour}
\affiliation{Department of Physics, University of Mohaghegh Ardabili, P.O. Box 179, Ardabil, Iran}

\author{J. Cheraghalizadeh}
\affiliation{Department of Physics, University of Mohaghegh Ardabili, P.O. Box 179, Ardabil, Iran}
\email{jafarcheraghalizadeh@gmail.com}

\author{H. Dashti-Naserabadi}
\affiliation{School of Physics, Korea Institute for Advanced Study, Seoul 130-722, South Korea}
\email{h.dashti82@gmail.com}

\begin{abstract}
The statistical properties of the carrier density profile of graphene in the ground state in the presence particle-particle interaction and random charged impurity in zero gate voltage has been recently obtained by Najafi \textit{et al.} (Phys. Rev E95, 032112 (2017)). The non-zero chemical potential ($\mu$) in gated graphene has non-trivial effects on electron-hole puddles, since it generates mass in the Dirac action and destroys the scaling behaviors of the effective Thomas-Fermi-Dirac theory. We provide detailed analysis on the resulting spatially inhomogeneous system in the framework of the Thomas-Fermi-Dirac theory for the Gaussian (white noise) disorder potential. We show that, the chemical potential in this system as a random surface, destroys the self-similarity, and the charge field is non-Gaussian. We find that the two-body correlation functions are factorized to two terms: a pure function of the chemical potential and a pure function of the distance. The spatial dependence of these correlation functions is double-logarithmic, e.g. the two-point density correlation $D_2(r,\mu)\propto \mu^2\exp\left[-\left(-a_D\ln\ln r^{\beta_D}\right)^{\alpha_D} \right]$ ($\alpha_D=1.82$, $\beta_D=0.263$ and $a_D=0.955$). The Fourier power spectrum function behaves like $\ln(S(q))=-\beta_S^{-a_S}\left(\ln q \right)^{a_S}+2\ln \mu$ ($a_S=3.0\pm 0.1$ and $\beta_S=2.08\pm 0.03$) in contrast to the ordinary Gaussian rough surfaces for which $a_S=1$ and $\beta_S=\frac{1}{2}(1+\alpha)^{-1}$, ($\alpha$ being the roughness exponent). The geometrical properties are however similar to the un-gated ($\mu=0$) case, with the exponents that are reported in the text.
\end{abstract}
\maketitle
\section{Introduction}
Graphene is a $2+1$ electron system which can be described at low energies by massless Dirac-Fermion model. Many studies on the unusual properties of graphene are still based on idealized models which neglect the effect of disorder and particle-particle interactions. The understanding of the origin and effects of extrinsic disorder, as well as interactions in graphene seems to be essential in understanding the experiments and also in designing graphene-based electronic devices. \\
The effect of the electron-electron interaction as well as the disorder in graphene is a long-standing problem~\cite{Mishchenko,HwangAdamSarma,HwangSarma}. The interplay of the electron-electron interaction and the disorder in graphene leads to some interesting phenomenon. An important phenomena whose explanation needs a method that treats these two in a same footing, is the appearance of strong carrier density inhomogeneity with density fluctuations much larger than the average density of the system for low densities, i.e. electron-hole puddles (EHPs)~\cite{Martin,HwangAdamSarma} which is believed to be responsible for the observed minimum conductivity of graphene~\cite{Rossi}. Around the zero gate voltage the transport is governed by the complex network of small random puddles with semimetal character, depending on the details of the charged impurity configuration in the sample. It has been proposed that such an inhomogeneity dominates the graphene physics at low ($\lesssim 10^{12}$ cm$^{-2}$) carrier densities \cite{Rossi} for which self-consistent Thomas-Fermi-Dirac (TFD) theory was employed to simulate the graphene charge profile on the SiO$_2$ substrate. EHPs were theoretically predicted by Hwang \textit{et al} \cite{HwangAdamSarma} and Adam \textit{et al} \cite{Adam} as the phase of low carrier density. The existence of these inhomogeneities, characterized by strong electron density fluctuations, were also confirmed in experiments in the vicinity of the Dirac point \cite{Martin,Rutter,Brar,ZhangBrar,Deshpande1,Martin2,Deshpande2,Ishigami,ChoFuhrer,Berezovsky1,Berezovsky2}. Recently an attempt concerning this point was made in which it was claimed that the contour lines in the graphene membranes are also conformal invariant \cite{herman2016}.

A substantial feature of experiments on graphene near the Dirac point is the formation of large (spanning) clusters of negative or positive charge densities. The presence of the spanning cluster in a system may be the fingerprint of the scale invariance and the self-affinity of the system. This leads to some scaling behaviors which is expected to present in scale-free systems \cite{Najafi1,Najafi2,Najafi3,Najafi4}. The presence of carrier charge self-similarity is an important question in graphene. The zero chemical potential has been analyzed recently by Najafi \textit{et al}~\cite{Najafi2017scale}, and is shown to be very different from the ordinary 2D electron gas. In the low density limit in which the charge fluctuation is maximal, using the Schramm-Lowewner evolution theory, it was shown that this system is conformal invariant and some critical exponents were reported~\cite{Najafi2017scale,Najafi2018interaction}. By analyzing the iso-charge lines of the system at the Dirac point it has been shown that the fractal dimension of the corresponding random loops is $D_f(\mu=0)=1.38\pm 0.02$. \\

There are increasing numerical and experimental evidences that the iso-height lines in the random fluctuating fields in $(2+1)-$ rough surfaces are scale invariant. The size distribution in these systems is characterized by a few scaling functions and scaling exponents \cite{kondevpre}. The graphene system, as a $2+1$ random system can be mapped to the rough surfaces ~\cite{Najafi2017scale,Najafi2018interaction}. By analyzing the contour lines of the electron-hole density, the authors showed that un-gated graphene ($\mu=0$) is non-Gaussian self-similar system. In the present paper we test numerically the Thomas-Fermi-Dirac theory for the finite-$\mu$ graphene system, and investigate the local and geometrical properties of the random charge of the system. The finite chemical potential breaks apparently the scale invariance of the system. We show that distribution of electrons and holes become different for finite $\mu$s and the two-point functions are factorized into two different pure functions: one a function of $\mu$ and the other a function of the spatial scale. The numerical fits of these functions show that their dependence are double-logarithmic with the distance. We also investigate the geometrical properties of the system and show that when the surface is cut in the vicinity of the mean density, the geometrical properties of the system is similar to the un-gated graphene. \\

The paper is organized as follows. In the next section we will introduce the TFD model. In the SEC.~\ref{RoughSurface} we will fix the notation and introduce different scaling behaviors and scaling exponents corresponding to the contour loop ensembles (CLE). In the SEC.~\ref{num} we will numerically measure the proposed scaling exponents for the disorder potential and the carrier density in graphene. In the final section, we summarize the obtained results and our conclusions.

\section{Ground state of Graphene}\label{groungstate}

In graphene the carrier density is controlled by the gate voltage $n=\kappa_SV_g/4\pi t$ in which $\kappa_S$ is the substrate dielectric constant and $t$ is its thickness and $V_g$ is the gate voltage. The experimental data show a strong dependence on $x\equiv n/n_i$ in which $n_i$ is the impurity density. In ordinary densities, the conductivity is linear function of $x$ and for very low $x$'s, it reaches a minimum of order $\sigma\sim e^2/h$ which is linked to the formation of EHP's. The first scanning probe experiment on exfoliated graphene on SiO$_2$ substrate were done by Ishigami \textit{et al} \cite{Ishigami} revealing its atomic structure and nano-scale morphology. Martin \textit{et al} used the scanning single-electron transistor (SET) to investigate the atomic structure and charge profile of exfoliated graphene close to the Dirac point. Interestingly a high electron density inhomogeneity, breaking up the density landscape in electron-hole puddles were observed in this experiment supporting the theoretical predictions of Adam \textit{et al} \cite{Adam} and Hwang \textit{et al} characterized by large scale electron density fluctuations \cite{HwangAdamSarma}. This strong fluctuations bring the system into a new phase with broken homogeneity in which some random electron and hole conducting puddles are created \cite{HwangAdamSarma}. \\
The source of disorder and its relevance in the electronic structure of graphene is also an important question to be addressed. The approximately linear dependence of conductivity on carrier density in graphene sheets \cite{Nomura,HwangAdamSarma} indicates that the remote Coulomb impurities are dominant disorder source in most graphene samples. The experimental observation that the spatial pattern of EHPs is not correlated with the topography of the graphene sheets (described in the previous subsection) is another evidence that the remote charges are the dominant disorder source \cite{Barlas}. The inclusion of Coulomb disorder in graphene in the absence of particle-particle interaction were studied by Fogler \textit{et al.} to investigate diffusive and ballistic transport in graphene $p-n$ junction \cite{Fogler}. The disorder in addition of being the main sources of the scattering has an additional effect; it locally shifts the Dirac point. It means that even at the zero gate voltage, the Fermi energy is moved to positive or negative values with respect to the charge neutrality (Dirac) point. The other sources of scattering are ripples \cite{NetoKim} and point defects (which is responsible for high-density saturation of conductivity \cite{HwangAdamSarma}) which are not considered in this paper.\\
The case of relevance is a slow (spatial) varying charge density system. An approach similar in sprit to the LDA-DFT is the Thomas-Fermi-Dirac theory which is valid only for the case $|\nabla_r n(\textbf{r})/n(\textbf{r})|\ll k_F(\textbf{r})$ in which $k_F(\textbf{r})$ is the Fermi wave number at position $\textbf{r}$. It has been shown that for the clean graphene in the low density regime $n\rightarrow 0$ the exchange potential goes to zero such as $V_x(n\rightarrow0)\varpropto -\text{sgn}(n)\sqrt{n}\ln |n|$ as well as the correlation potential, for which the proportionality constant will be introduced below \cite{Polini}. Using local density approximation one can prove that the total energy of the graphene for a disorder configuration and a density profile is \cite{SarmaRevModPhys}:
\begin{eqnarray}
\begin{split}
E=&\hbar v_F[\frac{2\sqrt{\pi}}{3}\int d^2r\text{sgn}(n)|n|^{\frac{3}{2}}\\
&+\frac{r_s}{2}\int d^2r\int d^2r^{\prime}\frac{n(\textbf{r})n(\textbf{r}^{\prime})}{|\textbf{r}-\textbf{r}^{\prime}|}\\
&+r_s\int d^2rV_{xc}[n(\textbf{r})]n(\textbf{r})+r_s\int d^2rV_D(\textbf{r})n(\textbf{r})\\
&-\frac{\mu}{\hbar v_F}\int d^2rn(\textbf{r})]
\end{split}
\end{eqnarray}
in which $v_F$ is the Fermi velocity, $r_s\equiv e^2/\hbar v_F\kappa_S$ is the dimensionless interaction coupling constant, $\mu$ is the chemical potential, $g=g_sg_v=4$ is the total spin and valley degeneracy. The exchange-correlation potential has been shown to be \cite{Polini}:
\begin{eqnarray}
V_{xc}=\frac{1}{4}\left[ 1-gr_s\zeta(gr_s)\right]\text{sgn}(n)\sqrt{\pi|n|}\ln\left(4k_c/\sqrt{4\pi |n|}\right)
\end{eqnarray}
in which $k_c$ is the momentum cut-off and $\zeta(y)=\frac{1}{2}\int_0^{\infty}\frac{dx}{(1+x^2)^2\left( \sqrt{1+x^2}+\pi y/8\right)}$.\\
For zero chemical potential $\mu=0$, the charged impurities are not screened in graphene~\cite{Polini}. For non-zero $\mu$s however, the screening effects become important and cannot be neglected. It has been shown that in the $q\rightarrow 0$ limit the Thomas-Fermi dielectric function becomes $\epsilon_{\text{TF}}\equiv \epsilon_{\text{RPA}}(q\rightarrow 0)=1+\frac{q_{\text{TF}}}{q}$ in which $q_{\text{TF}}=g_sg_ve^2/\kappa_S v_F$~\cite{hwang2007dielectric}.  Within this approximation, the potential of a charged impurity located at a distance $d$ from the substrate is $\tilde{v}(q)=\frac{2\pi e^2}{\kappa}\frac{e^{-qd}}{q+q_{\text{TF}}}$ which gives rise to the following form for the real space~\cite{adam2008boltzmann}:
\begin{eqnarray}
V(\textbf{r})=\frac{e^2}{\kappa_S}\frac{e^{-q_{\text{TF}} r}}{\sqrt{r^2+d^2}}
\label{Eq:potential}
\end{eqnarray}
in which $r\equiv |\textbf{r}|$. Therefore the potential of Eq.~\ref{VDis} should be replaced by the Yukawa potential in the finite $\mu$s. Therefore the remote Coulomb disorder potential is calculated by the relation:
\begin{eqnarray}
V_D(r)=\int d^2r^{\prime}\frac{\rho(\textbf{r}^{\prime})e^{-q_{\text{TF}} |\textbf{r}-\textbf{r}'|}}{\sqrt{|\textbf{r}-\textbf{r}^{\prime}|^2+d^2}}
\label{VDis}
\end{eqnarray}
in which $\rho(r)$ is the charged impurity density and $d$ is the distance between substrate and the graphene sheet. For the graphene on the SiO$_2$ substrate, $\kappa_S\simeq 2.5$, so that $r_s\simeq0.8$, $d\simeq 1$ nm, $k_c=1/a_0$ where $a_0$ is the graphene lattice constant $a_0\simeq 0.246$ nm corresponding to the energy cut-off $E_c\simeq 3$ eV. It is notable that in the above equations we have considered bare coulomb interactions for both impurity and Hartree terms. This is due to the absence of screening in low career densities, i.e. in the vicinity of the Dirac points. To obtain the equation governing $n(r)$ one can readily minimize the energy with respect to $n(r)$ which yields:
\begin{eqnarray}
\begin{split}
&\text{sgn}(n)\sqrt{|\pi n|}+\frac{r_s}{2}\int d^2 \textbf{r}^{\prime}\frac{n(\textbf{r}^{\prime})}{|\textbf{r}-\textbf{r}^{\prime}|}\\
&+r_sV_{xc}[n]+r_sV_D(\textbf{r})-\frac{\mu}{\hbar v_F}=0
\end{split}
\label{mainEQ}
\end{eqnarray}
which should be solved self-consistently. In this paper we consider the disorder to be white noise with Gaussian distribution $\left\langle \rho(\textbf{r})\right\rangle=0 $ and $\left\langle \rho(\textbf{r})\rho(\textbf{r}^{\prime})\right\rangle=(n_id)^2\delta^2(\textbf{r}-\textbf{r}^{\prime})$. Due to pure $1/r$ dependence of the Hartree and disorder terms, the convergence of the equation is slow.
\subsection{Scaling and the Probability Measure}
Let us now concentrate on the scaling properties of this equation ignoring $V_{xc}$. By zooming out the system, i.e. the transformation $\textbf{r}\rightarrow \lambda \textbf{r}$, we see that for the case $\mu=0$ the equation remains unchanged if we transform $n(\textbf{r})\rightarrow n(\lambda\textbf{r})=\lambda^{-2}n(\textbf{r})$ as expected from the spatial dimension of $n(\textbf{r})$. This is because of the fact that $V_D(\lambda\textbf{r})=\lambda^{-1}V_D(\textbf{r})$. This symmetry is very important, since it causes the system to be self-affine and may be violated for other choices of disorder. This scale-invariance in two dimensions leads to power-law behaviors and some exponents which are vital for surface characterization. It may also lead to conformal invariance of the system, and if independent of the type of the disorder, brings the graphene surface into a member of the minimal conformal series. The existence of $V_{xc}$ makes things difficult, since $V_{xc}(\textbf{r})\rightarrow V_{xc}(\lambda\textbf{r})=\lambda^{-1}\left(V_{xc}-\beta\text{sgn}(n)\sqrt{\pi|n|}\ln\lambda\right) $ in which $\beta\equiv\frac{1}{4}(1-gr_s\zeta(gr_s))$. Therefore the rescaled equation is:
\begin{eqnarray}
\begin{split}
&\xi(\lambda)\ \text{sgn}(n)\sqrt{|\pi n|}+\frac{r_s}{2}\int d^2 \textbf{r}^{\prime}\frac{n(\textbf{r}^{\prime})}{|\textbf{r}-\textbf{r}^{\prime}|}\\
&+r_sV_{xc}[n]+r_sV_D(\textbf{r})=0
\end{split}
\end{eqnarray}
in which $\xi(\lambda)\equiv 1-\beta r_s\ln\lambda$. Therefore the first term survive marginally in the infra-red limit and scale invariance is expected, even in the presence of $V_{xc}$. The above symmetry is simply an additional symmetry which limits the correlation functions to show power-law behaviors, but further details of the system needs exact or numerical solution. One of the most important quantities in random field analysis is the probability measure of charge density $P(n)$. It is believed that the probability measure of charge density in graphene is not Gaussian \cite{SarmaRevModPhys}. It has been shown that the one-particle probability measure for the mono-layer graphene in the low $r_s$ limit is:
\begin{eqnarray}
\begin{split}
P_n=A\exp\left[-\zeta\left(\text{sgn}(n)\sqrt{\pi |n|}-\frac{\mu}{\hbar Sv_F}\right)\right]
\end{split}
\label{PnZeroth}
\end{eqnarray}
in which $\zeta\equiv \frac{4\ln (L/a)}{dn_i^2r_s}\bar{n}$, $A$ is a normalization constant and $S$ is the area of the sample. In the above equation, the effects of disorder and Hartree interaction have been coded in $\zeta$. It is clear that a very weak interaction, has the same effect as a very weak disorder, i.e. in both cases $\zeta^{-1}\rightarrow \infty$ which results to very wide charge distribution and large charge fluctuations. The other important limit is $\mu\rightarrow 0$ which implies that $\zeta^{-1}\rightarrow \infty$ resulting to large scale density fluctuations. This is the point we emphasized in the previous sub-sections: at the Dirac point the density fluctuations grow unboundedly which drives the system into a new phase, i.e. formation of EHPs, consistent with other theoretical results \cite{Rossi}. In this limit the power-law behaviors become possible. The chemical potential has two main effects: it controls $\bar{n}$ and its existence in $P(n)$ controls the density fluctuations. Its effects are studied in the SEC.~\ref{num}. \\

\section{Graphene as a rough surface, scaling properties of CLE}\label{RoughSurface}
In the above arguments we mentioned that $n(x,y)$ is the important field in our analysis by which the energy of the system is obtained. Characterizing this field is very important in distinguishing the local phases of the system and the phase separation pattern in the system. To this end we analyze the local and global properties of the system. The first analysis on the contour loop ensemble (CLE) of this system has been done in \cite{Najafi2017scale} in which the Eq.~\ref{mainEQ} was solved for $\mu=0$. This is shown by the relation $n(\lambda \mathbf{r}) \stackrel{d}{=} \lambda ^{\alpha} n(\mathbf{r})$ in which  $\alpha$ is \textit{roughness} exponent or the \textit{Hurst} exponent and $\lambda$ is a scaling factor and the symbol $\stackrel{d}{=}$ means the equality of the distributions. For self-similar surfaces ($mu=0$ in our case) the density-density correlation function $D_2(r) \equiv \langle \left[ n(\mathbf{r}+\mathbf{r_0})-n(\mathbf{r_0}) \right]^2 \rangle$ behaves like $ |\mathbf{r}| ^{2\alpha_l}$ and also the total variance $W(L)\equiv \langle \left[ n(\mathbf{r}) - \bar{n} \right]^2 \rangle _L$ behaves like $ L^{2\alpha_g}$ where the parameter $\alpha_l$ is called the local roughness exponent, $\bar{n}=\langle n(\mathbf{r}) \rangle_L$, $\langle \dots \rangle _L$ means that the average is taken over $\mathbf{r}$ in a box of size $L$, and the parameter $\alpha_g$ is the global roughness exponent. Self-affine surfaces are mono-fractals just if $\alpha_g = \alpha_l = \alpha$ \cite{barabasi1995fractal}. For these systems the Fourier power spectrum (the second moment of $n_{\textbf{q}}$, which is itself the Fourier component of $n(\textbf{r})$) behave like:
\begin{equation}
S_{\textbf{q}}\equiv\left\langle \left| n_{\textbf{q}}\right|^2 \right\rangle_{\mu=0}\sim\left| \textbf{q}\right|^{-2(1+\alpha)}
\end{equation}
and also the distribution function of the density $P( V )$ is of the Gaussian form $\frac{1}{\sigma\sqrt{2\pi}}e^{-\frac{V^2}{2\sigma^2}}$ where $\sigma$ is the standard deviation. It has been shown that $\alpha_l^{\mu=0}=0.35\pm 0.03$ and $\alpha_g^{mu=0}=0.38\pm 0.03$~\cite{Najafi2017scale}. Another quantity whose moments distributions should be Gaussian is the local curvature which is defined (at position $\mathbf{r}$ and at scale $b$) as $C_b(\mathbf{r}) = \sum_{m=1}^M \left[ n(\mathbf{r}+ b\mathbf{e}_m) - n(\mathbf{r}) \right]$, in which the offset directions $\left\lbrace\mathbf{e}_1,\dots,\mathbf{e}_M\right\rbrace$ are a fixed set of vectors whose sum is zero, i.e. $\sum_{m=1}^M \mathbf{e}_m =0$. If the rough surface is Gaussian, then the distribution of the local curvature $P (C_b)$ is Gaussian and the first and all the other odd moments of $C_b$ manifestly vanish since the random field has up/down symmetry $n(\mathbf{r})\longleftrightarrow -n(\mathbf{r})$. Additionally, for Gaussian random fields we have $\frac{\langle C_b^4 \rangle }{\langle C_b^2 \rangle ^2} = 3$.\\
The self-similarity can also be addressed in geometrical quantities like the fractal dimension of loops and also the exponents of the distribution function of gyration radius and loop length. One can extract the loops from the iso-density lines of the profile $n(\mathbf{r})$ at the level set $n(\mathbf{r}) = n_0$ from which some non-intersecting loops result which come in many shapes and sizes are obtained. For scale-invariant random surfaces these geometrical objects show various power-law behaviors. The exponent of the distribution functions of loop lengths $l$ ($P(l)$) and the gyration radius of loops $r$ ($P(r)$) are of especial importance. The fractal dimension of loops for un-gated graphene has numerically determined $D_f^{\mu=0}=1.38\pm 0.02$ and the exponent of the distribution of loop length $\tau_l^{\mu=0}=2.30\pm 0.02$~\cite{Najafi2017scale}.\\
As stated in the introduction for non-zero $\mu$s the scale invariance is broken and $n(x,y)$ (as a random field ) is not self-similar. The characterization of this symmetry-breaking is important in determining the system transport, e.g. the density-density correlation function which is important in determining the dielectric function. In the next section we address the deformation of various functions in the non-zero chemical potential limit.

\section{Numerical results and discussion}\label{num}

In this section we present the numerical results. This section has been divided to three sub-sections. In the next sub-section we present the results for the local quantities, i.e. the two-body (density-density) correlation function $D_2(r,\mu)$, the total variance $W_L(L,\mu)$ and the Fourier power spectrum $S_q$. In SEC.~\ref{higher} the higher moments are analyzed. The fractal dimension of loops and the distribution functions of the geometrical observables are calculated in SEC.\ref{Geo}.

\subsection{local exponents}\label{local}

\begin{figure*}
	\begin{subfigure}{0.45\textwidth}\includegraphics[width=\textwidth]{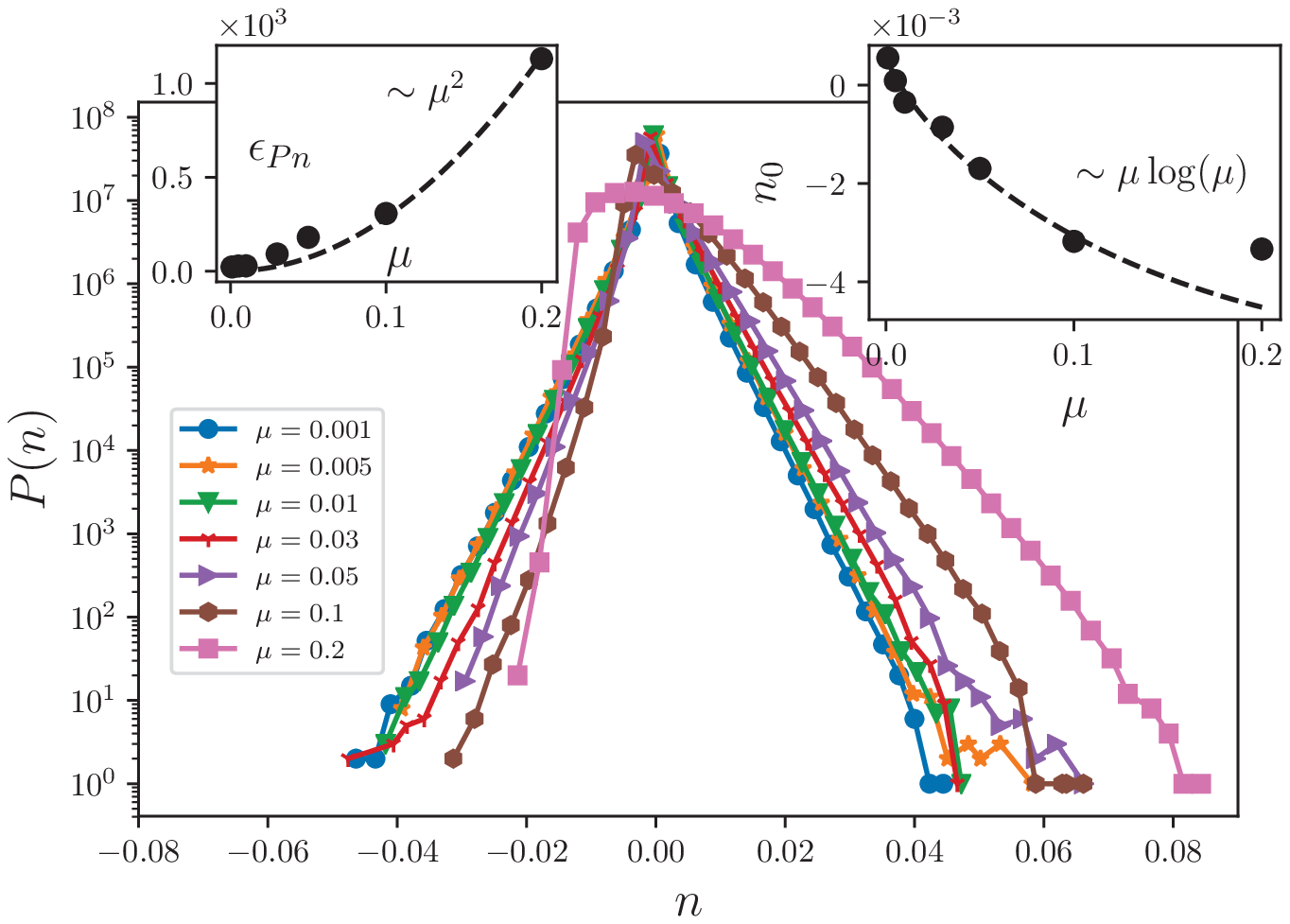}
		\caption{}
		\label{fig:pdf_H}
	\end{subfigure}
	\begin{subfigure}{0.45\textwidth}\includegraphics[width=\textwidth]{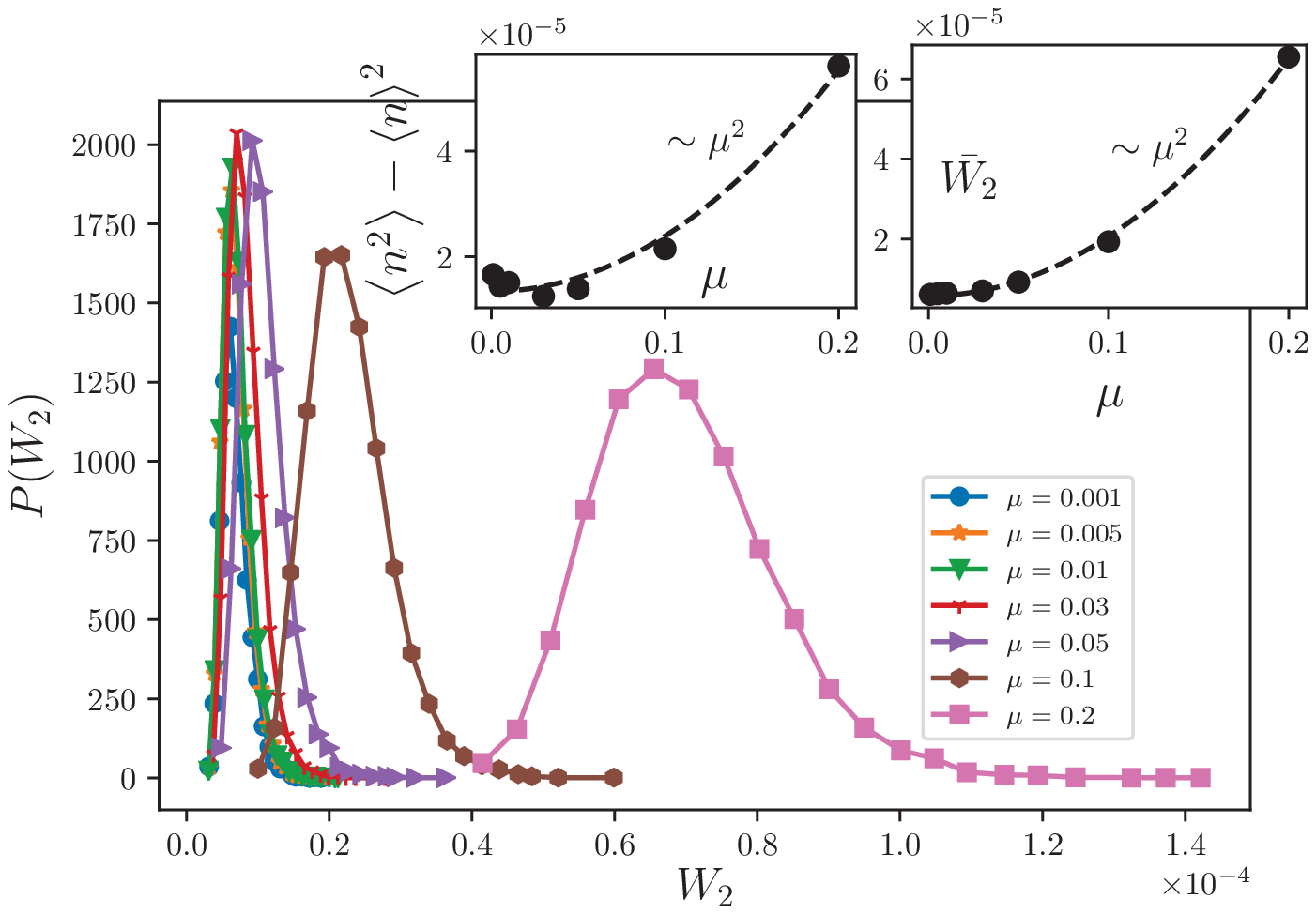}
		\caption{}
		\label{fig:pdf_W2}
	\end{subfigure}
	\caption{(Color Online) (a) The semi-log plot of the distribution function of density $P(n)$. Left inset: $\epsilon_{Pn}(\mu)$ which is defined as the (absolute value of the) difference between the slopes of two sides. Right inset: the position of the peak of the distribution function $n_0(\mu)$. (b) The distribution function of the roughness function $P(W_2)$ for the boxes of size $l=100$. Inset:  the position of the peak in terms of $\mu$.}
	\label{fig:distribution}
\end{figure*}

The chemical potential tunes the average density of any condensed matter system. When $\mu=0$ this average is zero, showing that we are right at the Dirac point. For non-zero values, the Fermi surface moves above or under the Dirac point and the system acquires non-zero average density. This can be seen in Fig.~\ref{fig:distribution} in which the distribution of $n$ has been shown. It is known that this function is non-Gaussian for graphene~\cite{Najafi2017scale} that is evident in this figure. This figure reveals that the logarithm of this function is linear in $n$ with two non-equal ($\mu$-dependent) slopes. For small $\mu$ values this function behaves like the following relation:
\begin{equation}
P_{\mu}(n)\propto
\begin{cases}  \exp\left[ -a_R \left( n-n_0(\mu)\right)\right] & n \geq n_0(\mu) \\
\exp\left[ a_L \left( n-n_0(\mu)\right)\right] & n < n_0(\mu)
\end{cases}
\label{Eq:Pn}
\end{equation}
in which $a_R$ and $a_L$ are the mentioned slopes (that are equal only for $\mu=0$) and $n_0(\mu)$ is the density at which the distribution function shows a peak ($n_0(\mu=0)=0$) and the function becomes singular. This function differs from the Eq.~\ref{PnZeroth} which has been calculated for low densities and low interactions and disorder strengths. In fact the form~\ref{Eq:Pn} is not true for very low densities, and $\log P(n)$ varies with $\sqrt{n}$ in accordance with the Eq.~\ref{PnZeroth}. We find form this figure that the asymmetry parameter $\epsilon_{Pn}(\mu)\equiv \left| a_L(\mu)-a_R(\mu)\right| $ increases with $\mu$ in a power-law fashion $\epsilon_{Pn} \sim \mu^2$, and $n_0(\mu)$ behaves like $\mu\ln\mu$ as depicted in Fig.~\ref{fig:distribution} (the left side and right side insets respectively). The asymmetry of $P_{\mu}(n)$ shows that the dynamics of electrons and holes in gated graphene is not the same as expected. Also it is seen that the graphs become wider for larger $\mu$ values showing that the density fluctuations (which is proportional to the system compressibility $\kappa_{\mu}$) increase by increasing $\mu$. Figure~\ref{fig:pdf_W2} shows the distribution of the total variance of density $P(W_2)$ (or roughness function in the rough surface language) for boxes with linear size $l=100$. We see that the peak of the distribution (which is here equal to the average value $\bar{W}_2$) scales with the square of the chemical potential, i.e. $\bar{W}_2\sim \mu^2$. This reveals that the density fluctuations increase with $\mu$ that is compatible with the widening of $P(n)$ shown above. To see this, we have plotted $(\delta n)^2\equiv \left\langle n^2\right\rangle - \left\langle n\right\rangle^2 $ in the left inset in this figure. It is evident that $\bar{W_2}$ and $(\delta n)^2$ have the same behaviors. The widening of the distribution function can be inferred from the Eq.~\ref{PnZeroth} in which large $\mu$ value is compensated by large density fluctuations. \\ 

\begin{figure*}
	\begin{subfigure}{0.45\textwidth}\includegraphics[width=\textwidth]{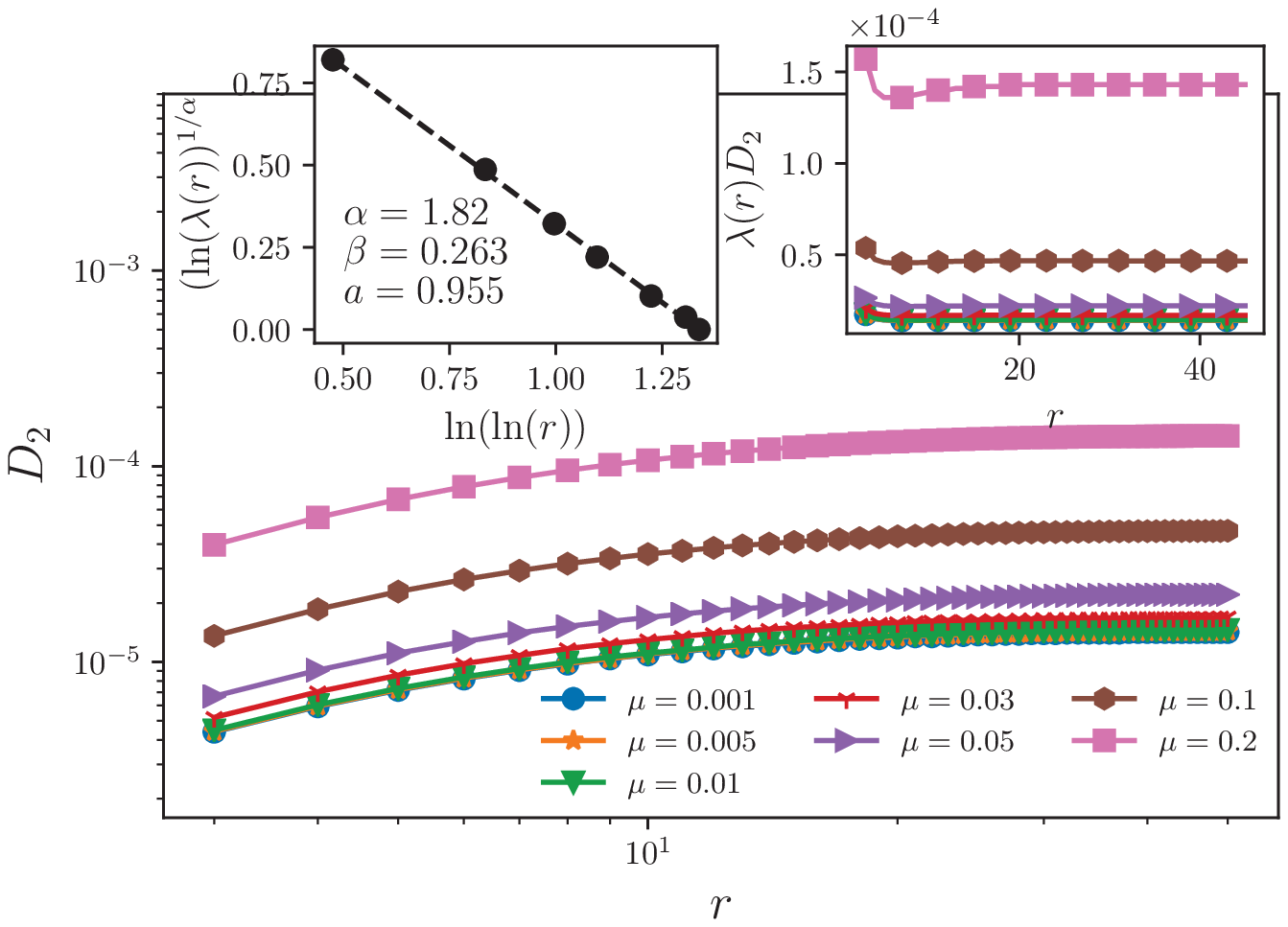}
		\caption{}
		\label{fig:D2_r}
	\end{subfigure}
	\begin{subfigure}{0.45\textwidth}\includegraphics[width=\textwidth]{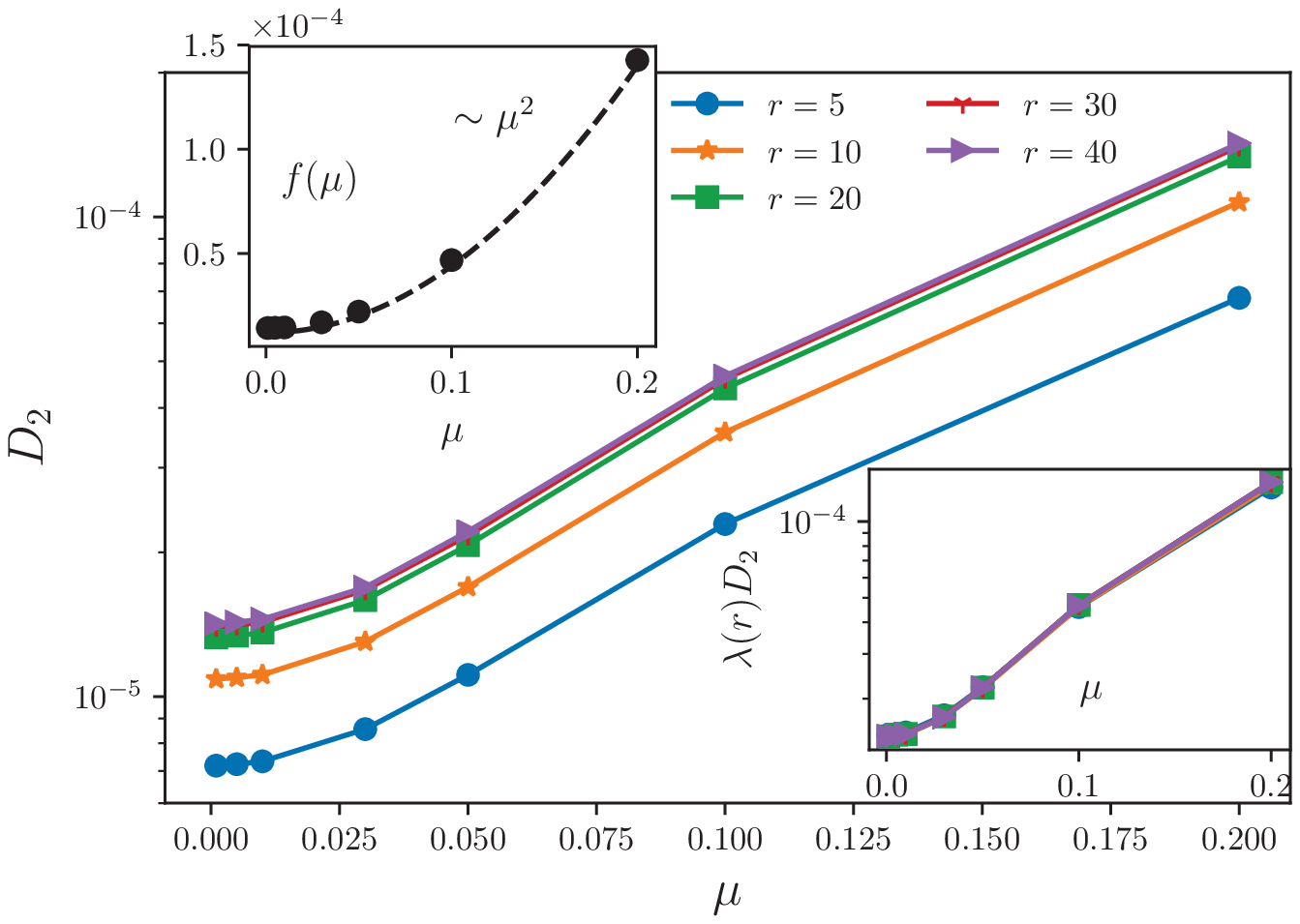}
		\caption{}
		\label{fig:D2_mu}
	\end{subfigure}
	\begin{subfigure}{0.45\textwidth}\includegraphics[width=\textwidth]{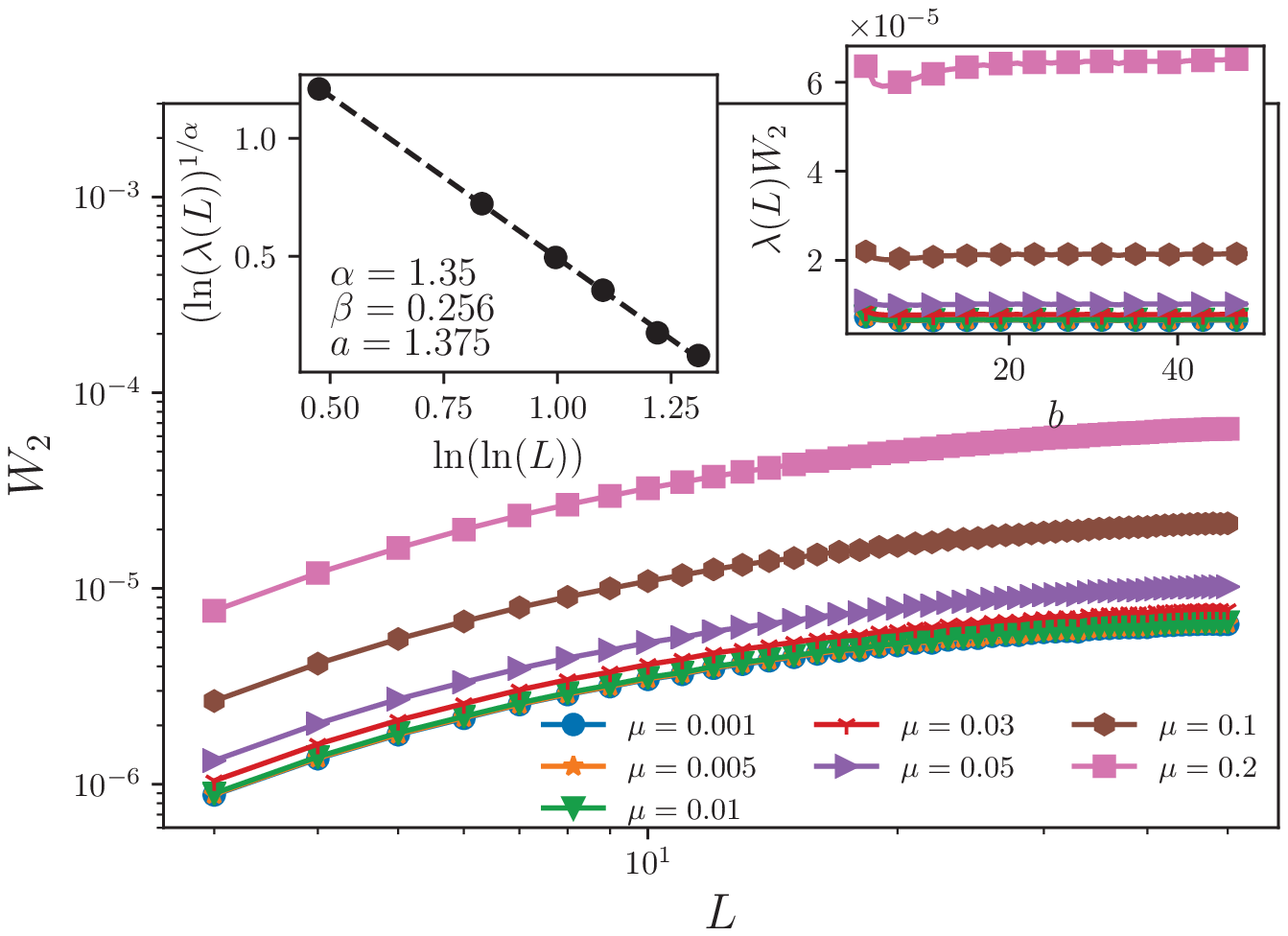}
		\caption{}
		\label{fig:Wb_2}
	\end{subfigure}
	\begin{subfigure}{0.45\textwidth}\includegraphics[width=\textwidth]{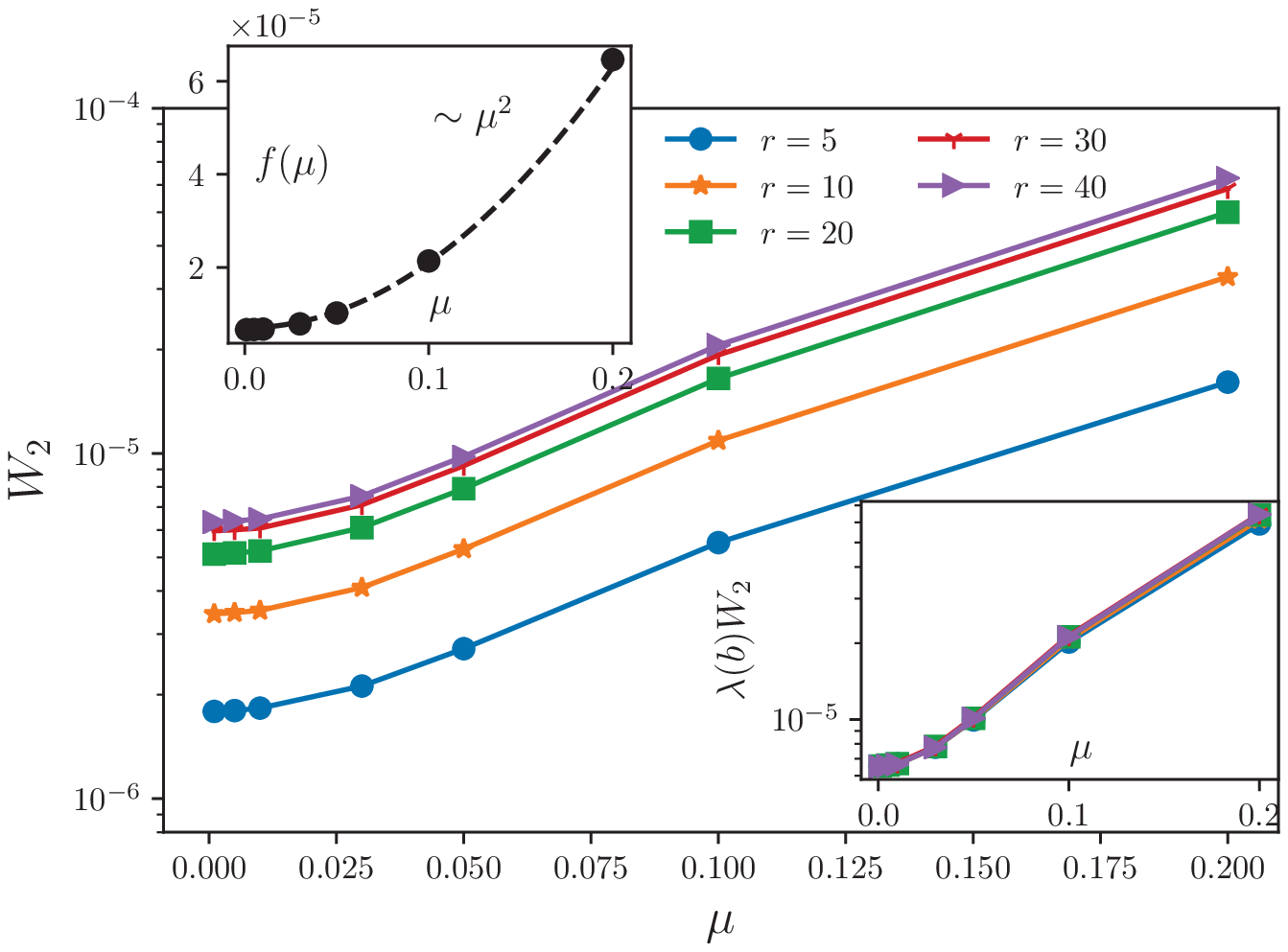}
		\caption{}
		\label{fig:W2_mu}
	\end{subfigure}
	\begin{subfigure}{0.45\textwidth}\includegraphics[width=\textwidth]{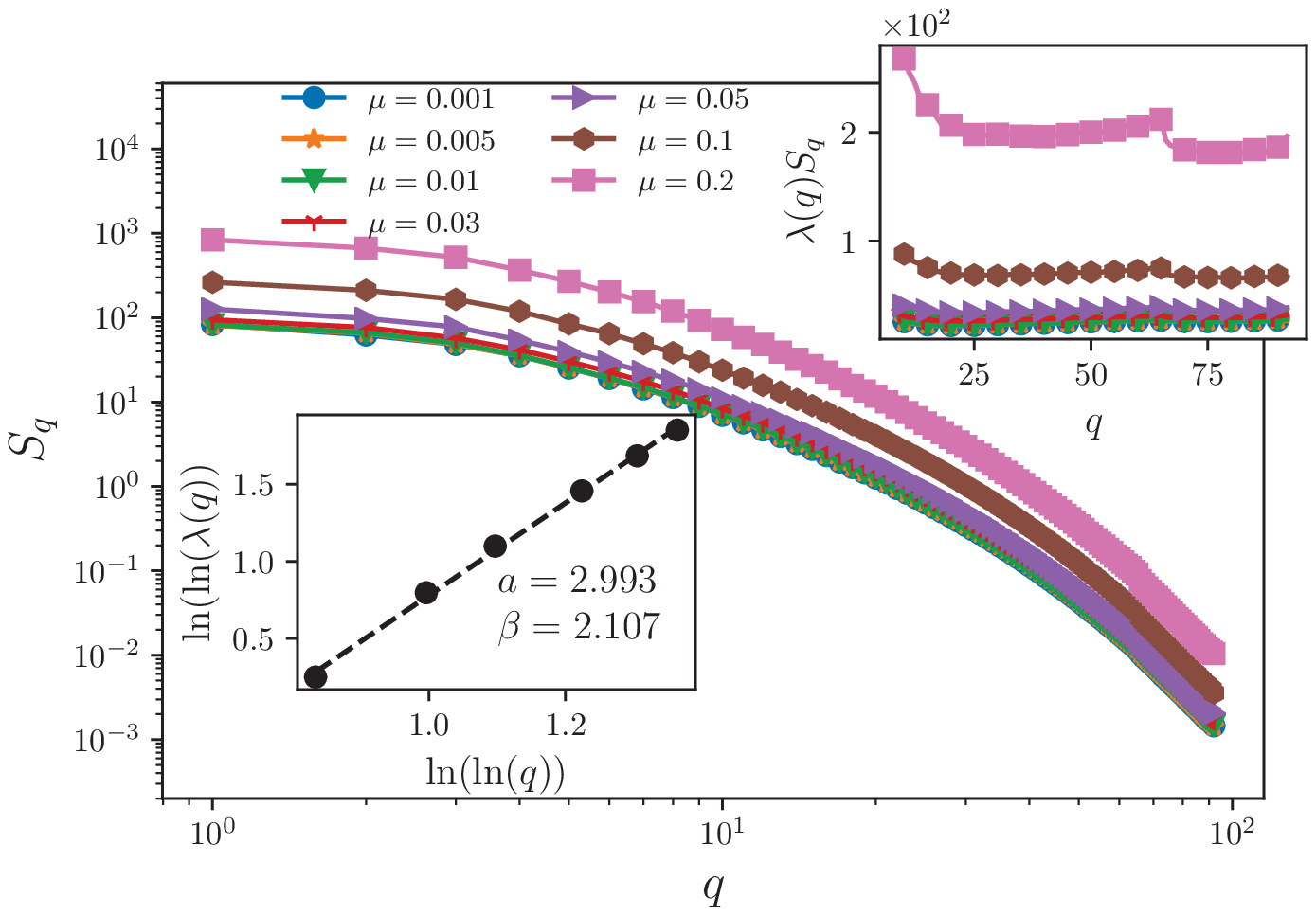}
		\caption{}
		\label{fig:Sq}
	\end{subfigure}
	\begin{subfigure}{0.45\textwidth}\includegraphics[width=\textwidth]{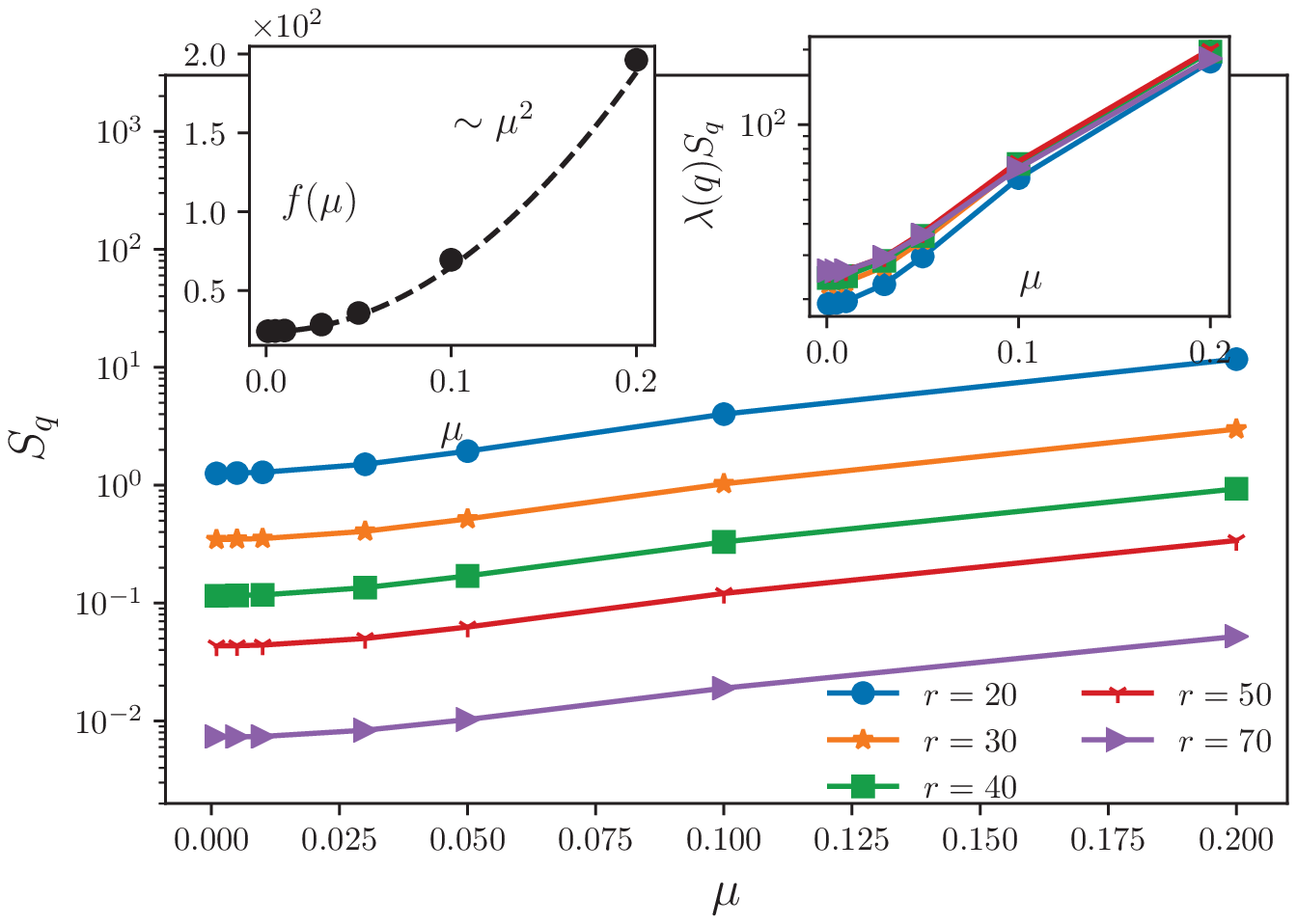}
		\caption{}
		\label{fig:Sq-mu}
	\end{subfigure}
	\caption{(Color Online) $D_2(r,\mu)=\frac{f_D(\mu)}{\lambda(r)}$ in terms of (a) $r$ and (b) $\mu$. The fitting is done for $\lambda(r)=\exp\left[\left(-a_D\ln\ln r^{\beta_D}\right)^{\alpha_D} \right] $ with the parameters $\alpha_D=1.82$, $\beta_D=0.263$, $a_D=0.955$, and $f_D(\mu)\sim \mu^2$. $W_2(r,\mu)=\frac{f_W(\mu)}{\lambda(L)}$ in terms of (c) $L$ and (d) $\mu$. The fitting is done for $\lambda(L)=\exp\left[\left(-a_W\ln\ln L^{\beta_W}\right)^{\alpha_W} \right] $ with the parameters $\alpha_W=1.35\pm 0.01$, $\beta_W=0.26\pm 0.01$, $a_W=1.375\pm 0.005$, and $f_W(\mu)\sim \mu^2$. $S_q=\frac{f_q(\mu)}{\lambda(q)}$ in terms of (e) $q$ and (f) $\mu$ with $\lambda(q)\exp \left[\left( \ln(q)/\beta_S\right)^{a_S} \right]$, $a_S=3.0\pm 0.1$, $\beta_S=2.1\pm 0.03$ and $f_S(\mu)\sim \mu^2$. }
	\label{fig:local}
\end{figure*}

Now let us consider the two point correlation functions which have been analyzed in Fig.~\ref{fig:local}, i.e. $D_2(r,\mu)$, $W_2(L,\mu)$ and $S_q(\mu)$. Interestingly we have observed that all multi-point functions considered in this paper are factorized to two parts: one part depending only on $\mu$ and the other a pure function of the other variable, e.g. $D_2(r,\mu)=\frac{f_D(\mu)}{\lambda(r)}$ (see the right hand inset of Fig.~\ref{fig:D2_r}). For this function, our analysis reveals that $f(\mu)\sim \mu^2$ and $\lambda(r)$ is best fitted by (the left inset of Fig.~\ref{fig:D2_r}):
\begin{equation}
\lambda(r)=\exp\left[\left(-a_D\ln\ln r^{\beta_D}\right)^{\alpha_D} \right]
\label{lambda}
\end{equation}
in which $\alpha_D=1.82\pm 0.01$, $\beta_D=0.26\pm 0.01$ and $a_D=0.96\pm 0.01$. This function, along the relevant quantities have been sketched in Fig~\ref{fig:D2_mu} in terms of $\mu$. The quadratic form of $f(\mu)$ has been shown in the insets. Therefore the two point function $D_2$ is not power-law in $r$, instead it behaves double logarithmic. This behavior is independent of $\mu$, which appears as a multiplicative constant, i.e. $f(\mu)$.\\
The same behavior is seen for $W_2(L)$ (in which $L$ is the box in which the variance (roughness) has been calculated, see SEC.~\ref{RoughSurface}) as is evident in Figs.~\ref{fig:Wb_2} (in terms of $L$) and \ref{fig:W2_mu} (in terms of $\mu$). In this figures $\lambda(L)$ is best fitted by the same relation as the Eq.~\ref{lambda}, with the parameters $\alpha_W=1.35\pm 0.01$, $\beta_W=0.26\pm 0.01$ and $a_W=1.375\pm 0.005$. The fact that $D_2$ and $W_2$ show the same $r$ and $\mu$ dependences is not surprising since the role of the spatial extent of the boxes in which the total variance is calculated is the same as the role of $r$ in the density-density correlation function. The example is the equality of the global and local roughness exponents in the scale-invariant rough surfaces.\\
A completely different behavior is seen for $S_q$. Our observations reveal that this function shows the following form:
\begin{equation}
\ln S_q=-\left(\beta_S^{-1}\ln q\right)^{a_S}+\ln f_{S}(\mu)
\label{Eq:Sq}
\end{equation}
in which $a_S=3.0\pm 0.1$, $\beta_S=2.1\pm 0.03$. Equivalently one finds that $S_q=\frac{f_{S}(\mu)}{\lambda(q)}$ in which $\lambda(q)=\exp\left[ \left( \ln(q)/\beta_S\right)^{a_S}\right]$. This relation shows that $\ln S_q$ for the graphene varies with third power of $\ln q$. This for should be compared with the same expression for the scale-invariant rough surface:
\begin{equation}
\ln S_q=-2\left(1+\alpha\right)\ln q + \text{const.}
\label{Eq:SqRough}
\end{equation}
Therefore, apart from the proportionality constant, the main difference of the $S_q$ of graphene and the ordinary rough surfaces is that the logarithm of the former depends on the third power of $\ln q$, whereas the latter is liner. The graphs for $S_q$ have been shown in Figs.~\ref{fig:Sq} and \ref{fig:Sq-mu}. It is evident from the Fig.~\ref{fig:Sq-mu} that $f_S(\mu)\sim \mu^2$ just like the functions $D_2$ and $W_2$. This is not surprising since $S_q$ is related to the Fourier transformation of $D_2$ and consequently with the same $\mu$ dependence.

\begin{figure*}
	\begin{subfigure}{0.45\textwidth}\includegraphics[width=\textwidth]{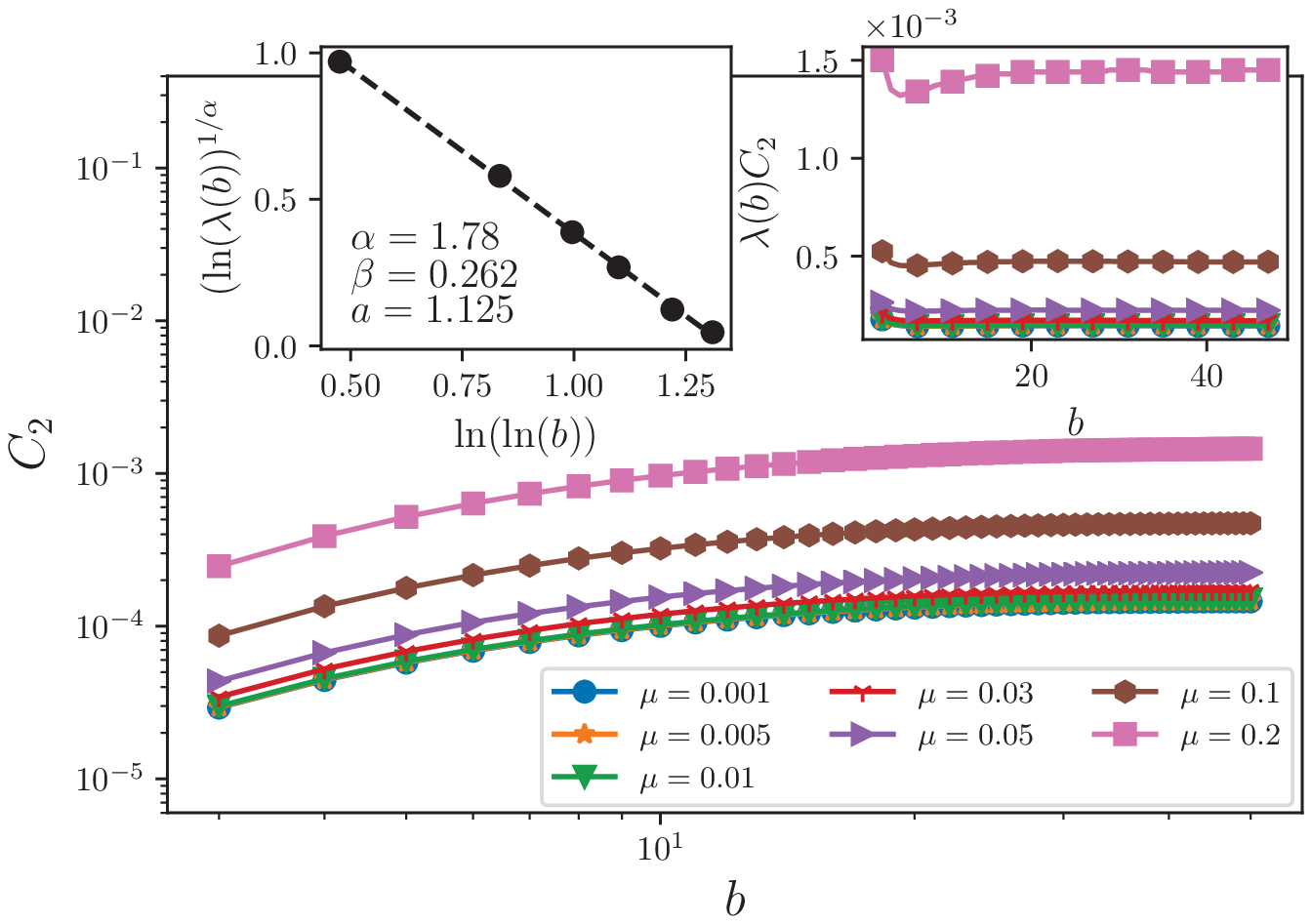}
		\caption{}
		\label{fig:Cb_2}
	\end{subfigure}
	\begin{subfigure}{0.45\textwidth}\includegraphics[width=\textwidth]{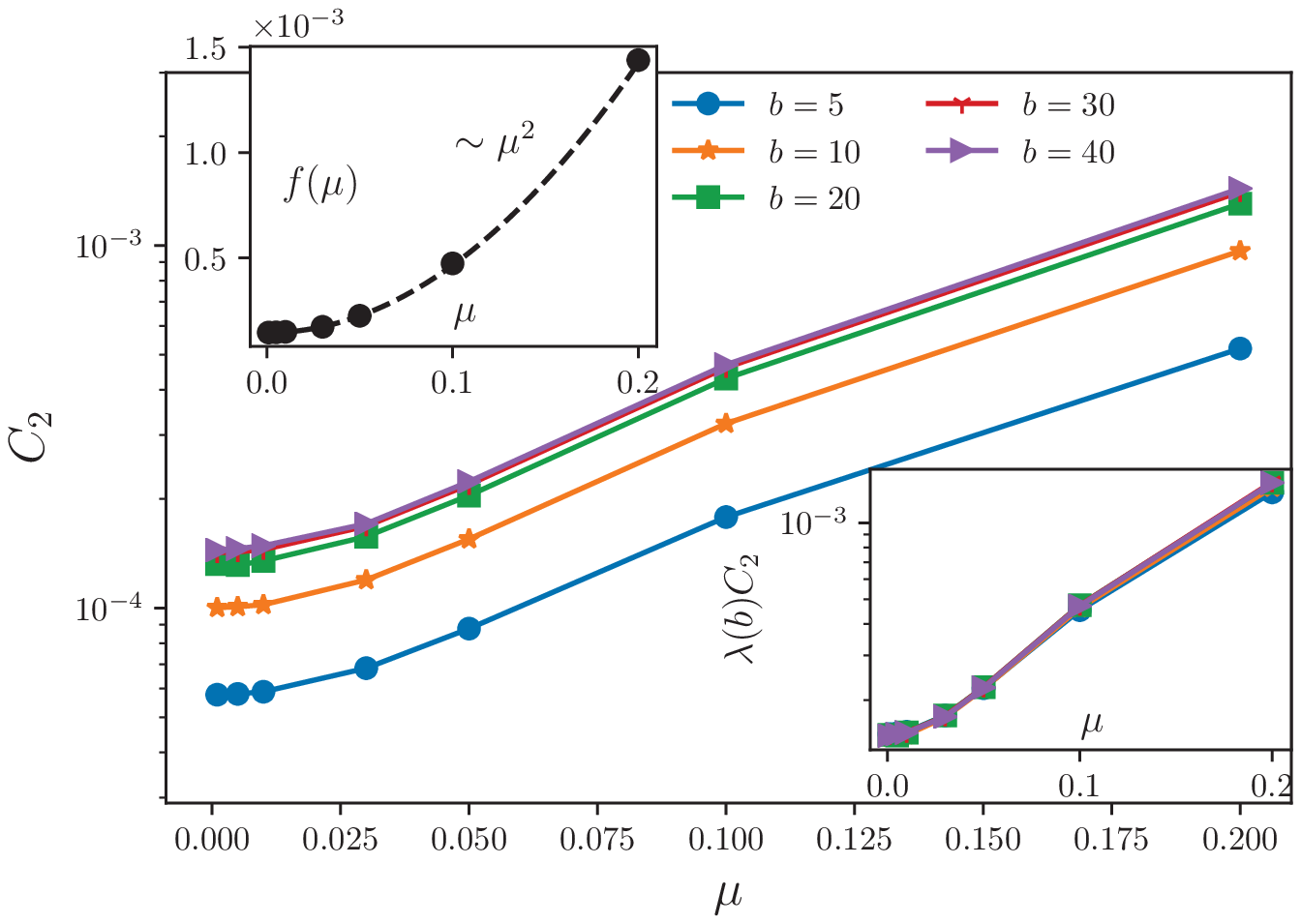}
		\caption{}
		\label{fig:C2_mu}
	\end{subfigure}
	\begin{subfigure}{0.45\textwidth}\includegraphics[width=\textwidth]{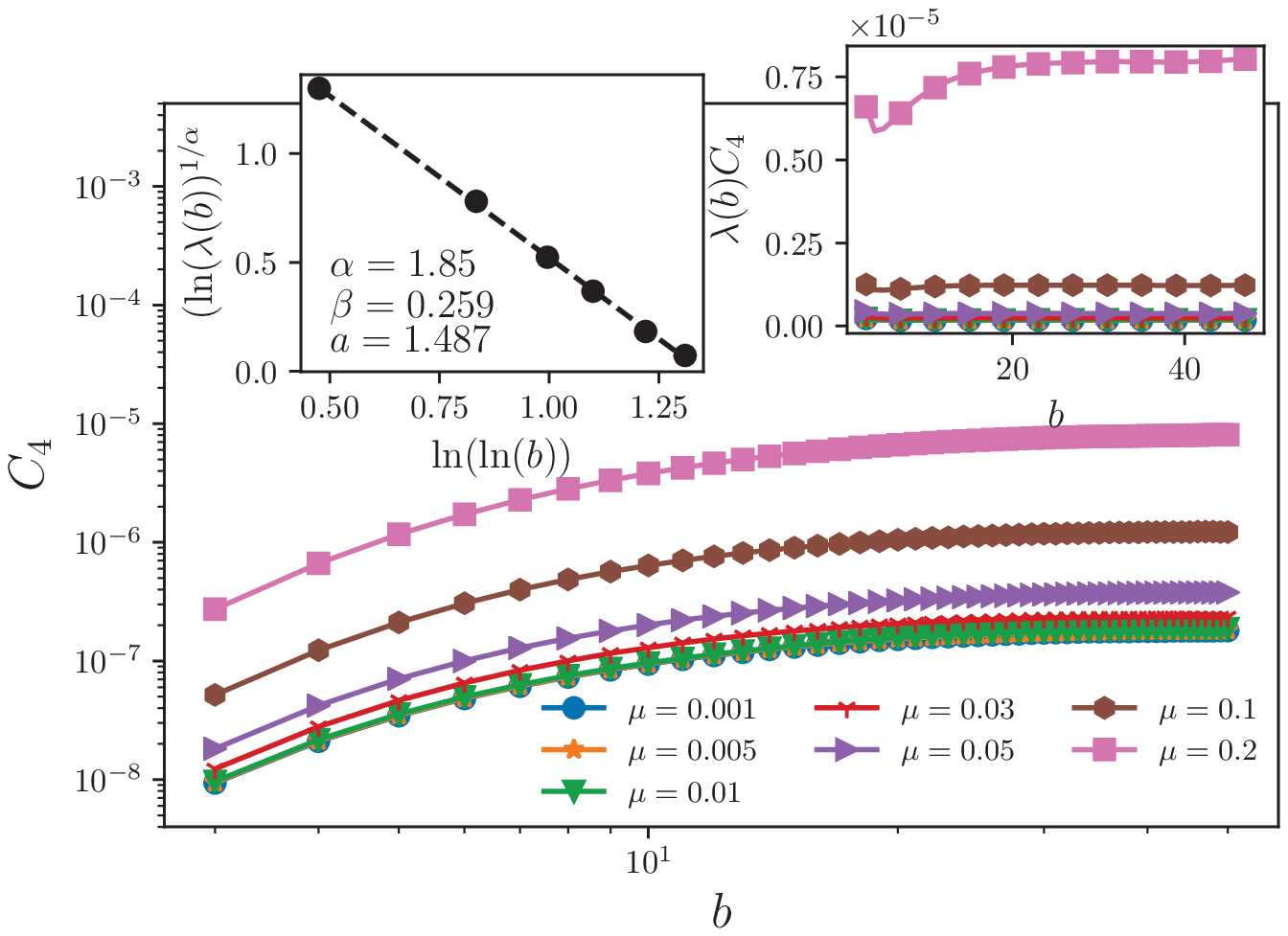}
		\caption{}
		\label{fig:Cb_4}
	\end{subfigure}
	\begin{subfigure}{0.45\textwidth}\includegraphics[width=\textwidth]{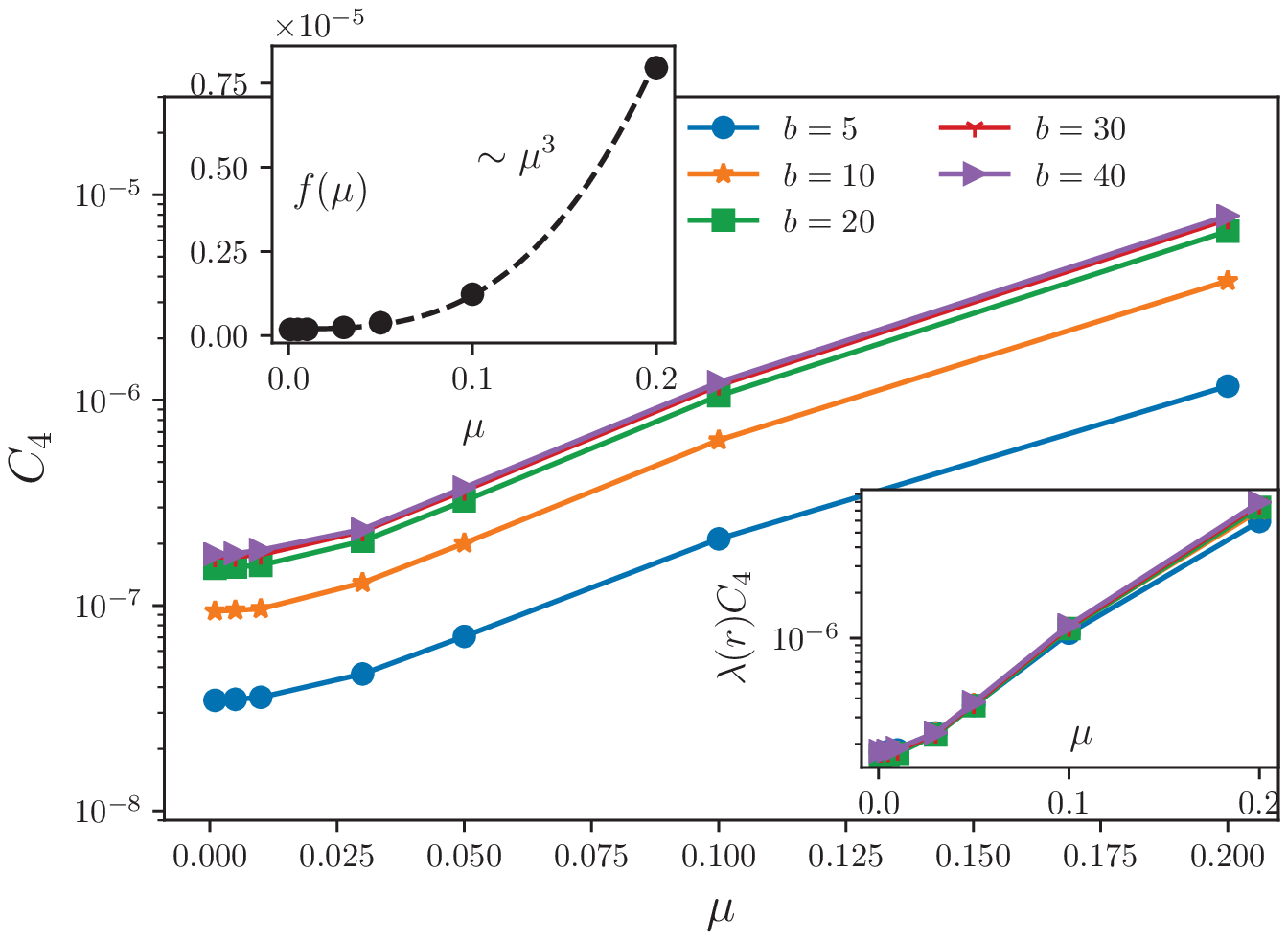}
		\caption{}
		\label{fig:C4_mu}
	\end{subfigure}
	\caption{(Color Online) $C_2(b,\mu)$ in terms of (a) $b$ (b) $\mu$, for which the fitting is done for $D_2(b,\mu)=\frac{f_C(\mu)}{\lambda(b)}$. The fitting parameters are $\lambda(b)\equiv \exp\left[\left(-a_C\ln\ln r^{\beta_C}\right)^{\alpha_C} \right]$, $\alpha_C=1.35\pm 0.02$, $\beta_C=0.25\pm 0.01$, $a_C=1.36\pm 0.01$ and $f_C(\mu)\sim\mu^2$. The same analysis for $C_4(b,\mu)$ in terms of (c) $b$ and (d) $\mu$, with the parameters $\alpha_C=1.85\pm 0.02$, $\beta_C=0.26\pm 0.01$, $a_C=1.86\pm 0.01$ and $f_C(\mu)\sim\mu^3$. }
	\label{fig:Cb}
\end{figure*}

The same features have been observed for $C_2(b)$ and $C_4(b)$ in which $C_n(b)\equiv \left\langle C_b^n\right\rangle$. These quantities have been shown in Fig.~\ref{fig:Cb}, in which $\alpha$, $\beta$ and $a$ have been reported for each case separately. The dependence on $b$ is just like Eq.~\ref{lambda} (with $r$ replaced by $b$). Also the dependence of $f_C(\mu)$ is power-law with the exponent $2$ (see fig.~\ref{fig:C2_mu}). This exponent is $3$ for $C_4$. This demonstrates that the distribution of $C_2$ and $C_4$ is not Gaussian and shows again that their dependence on $b$ is double-logarithmic.

\subsection{Higher order moments, non-Gaussian surface}\label{higher}

\begin{figure*}
	\begin{subfigure}{0.45\textwidth}\includegraphics[width=\textwidth]{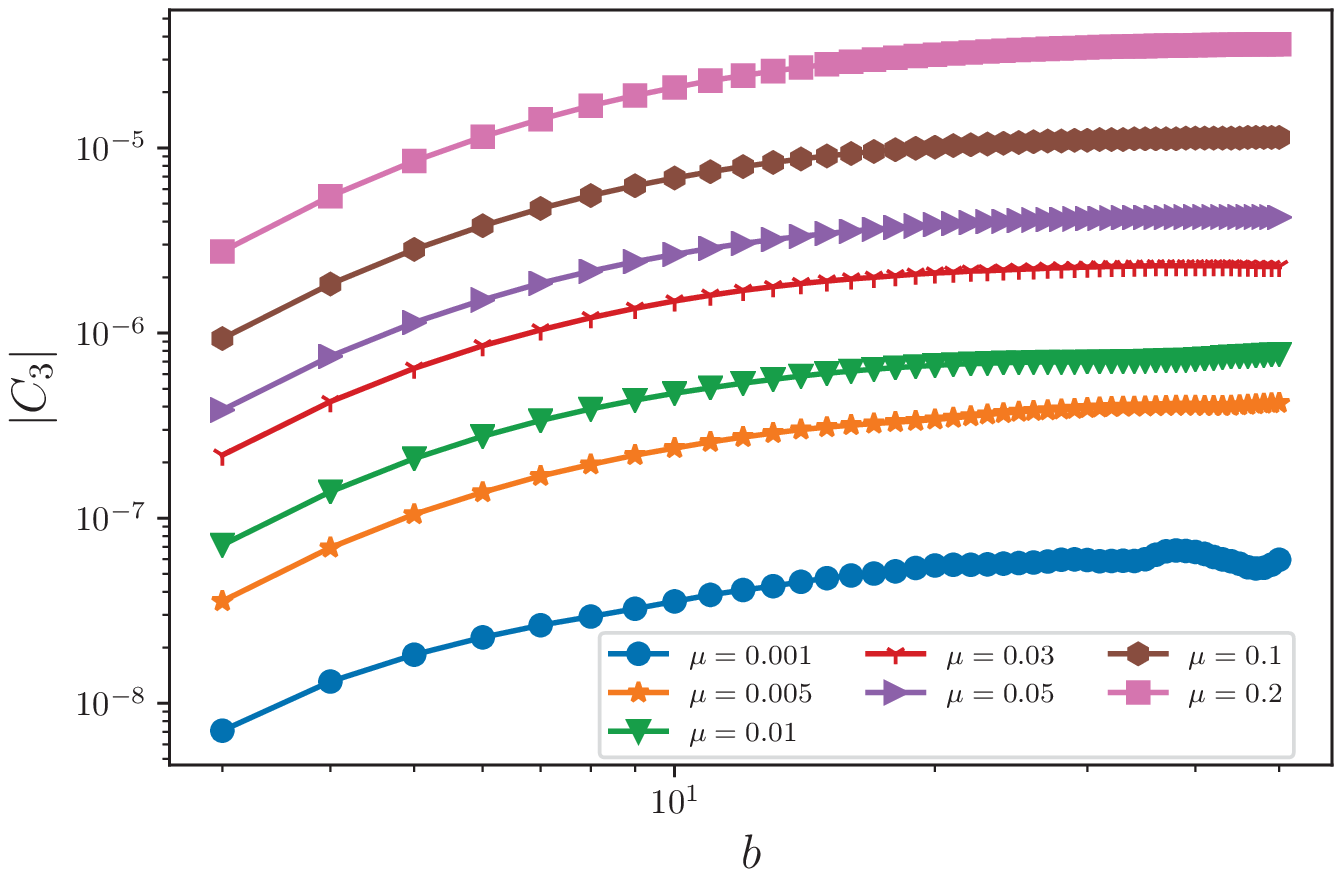}
		\caption{}
		\label{fig:Cb_3}
	\end{subfigure}
	\begin{subfigure}{0.45\textwidth}\includegraphics[width=\textwidth]{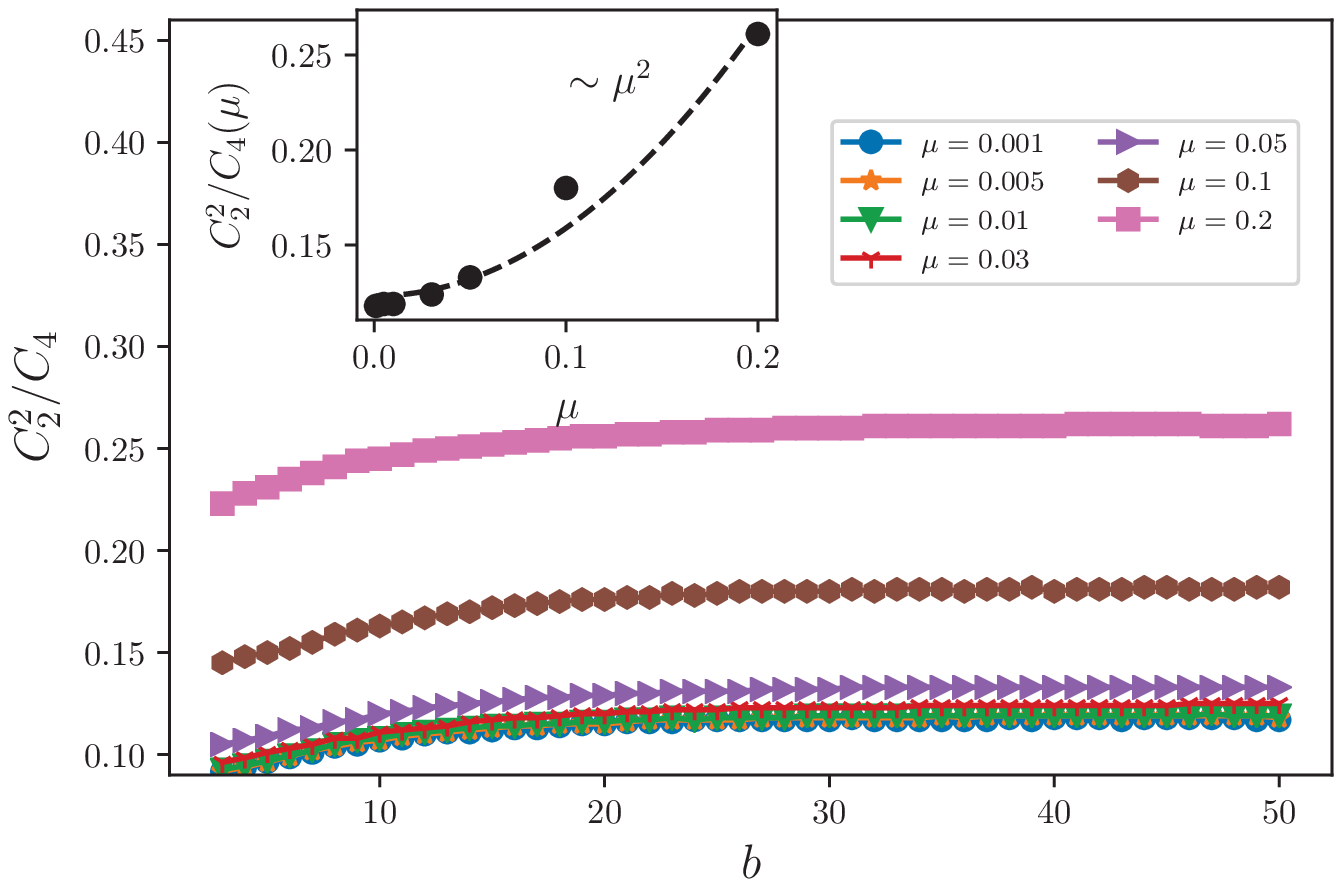}
		\caption{}
		\label{fig:C2_C4}
	\end{subfigure}
	\caption{(Color Online) The non-Gaussian parameters (a) $C_3(b,\mu)$ and $\left(C_2(b,\mu)\right)^2/C_4(b,\mu)$. The latter changes with the second power of $\mu$ which firms that the system is non-Gaussian.}
	\label{fig:Higher}
\end{figure*}

An important check for the systems which are mapped to the rough surfaces is the Gaussian/non-Gaussian behaviors. It is known that graphene is non-Gaussian rough surface, even at the Dirac point~\cite{Najafi2017scale}. However the exact characterization of this non-Gaussian rough surface needs some critical investigation on other variables, like $C_2(b,\mu)=\left\langle C_b^2\right\rangle$ (see the definition of $C_b(\textbf{r})$ in SEC~\ref{RoughSurface}), and the higher moments of density, especially the odd powers like $D_3(r,\mu)=\left\langle \left( n(\textbf{r}_0+\textbf{r})n(\textbf{r}_0)\right)^3 \right\rangle $ and $C_3(b,\mu)=\left\langle \left( C_b\right)^3 \right\rangle $, etc., which are expected to be zero for scale-invariant symmetric rough surfaces~\cite{kondevpre}. Figs.~\ref{fig:Higher} shows these functions.  

We see from Fig~\ref{fig:Cb_3} that $C_3$ does not vanish, and increases with $b$. Also we know that $\frac{C_2^2}{C_4}$ that, for a scale-invariant rough surface should be constant (for all $b$s) equal to $\frac{1}{3}$. Fig~\ref{fig:C2_C4} shows that it is not the case, and this function has a non-trivial increasing behavior in terms of $b$. All of these show that the system is non-Gaussian.

\subsection{Geometrical quantities}\label{Geo}

\begin{figure*}
	\begin{subfigure}{0.3\textwidth}\includegraphics[width=\textwidth]{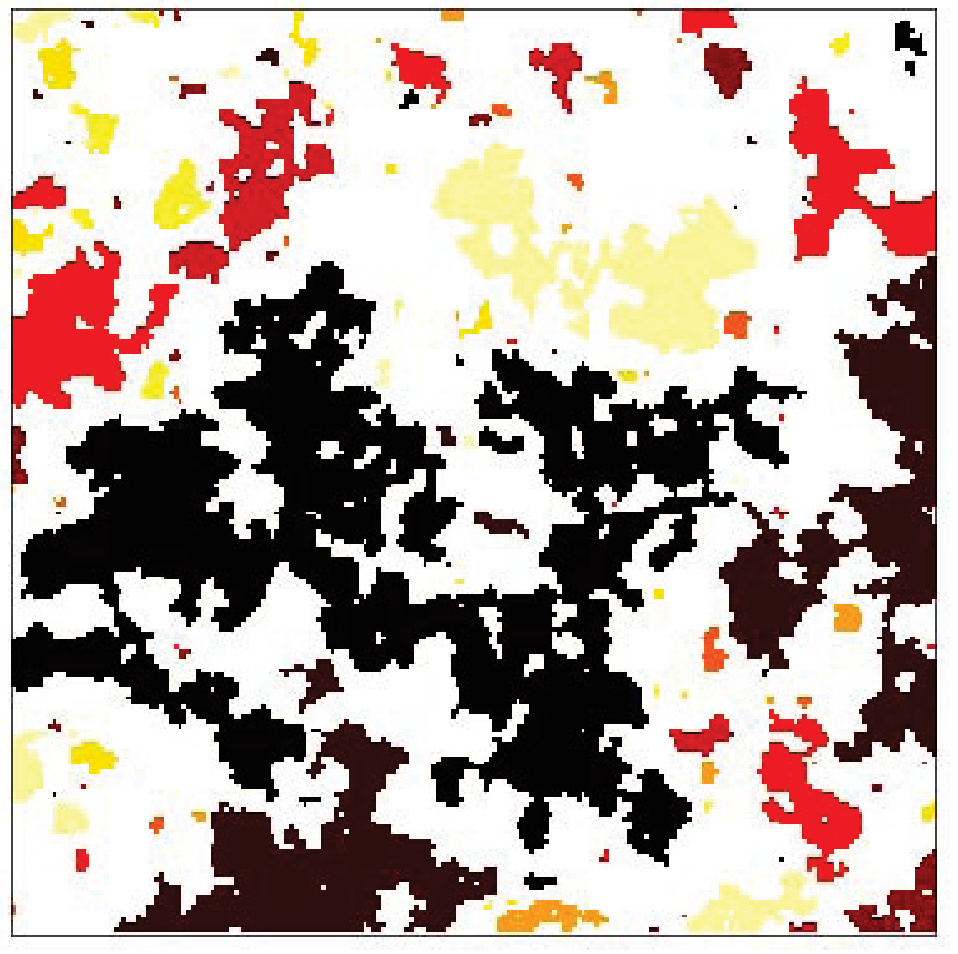}
		\caption{}
		\label{fig:mu001}
	\end{subfigure}
	\begin{subfigure}{0.3\textwidth}\includegraphics[width=\textwidth]{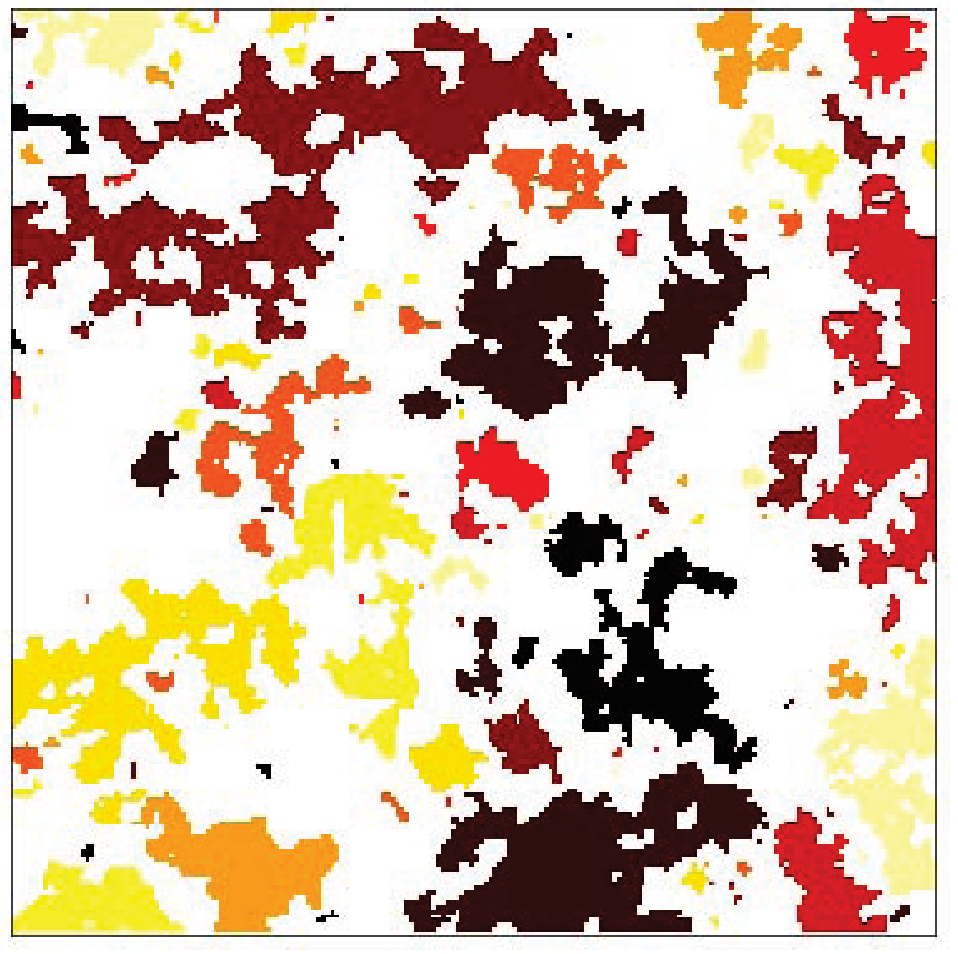}
		\caption{}
		\label{fig:mu005}
	\end{subfigure}
	\begin{subfigure}{0.3\textwidth}\includegraphics[width=\textwidth]{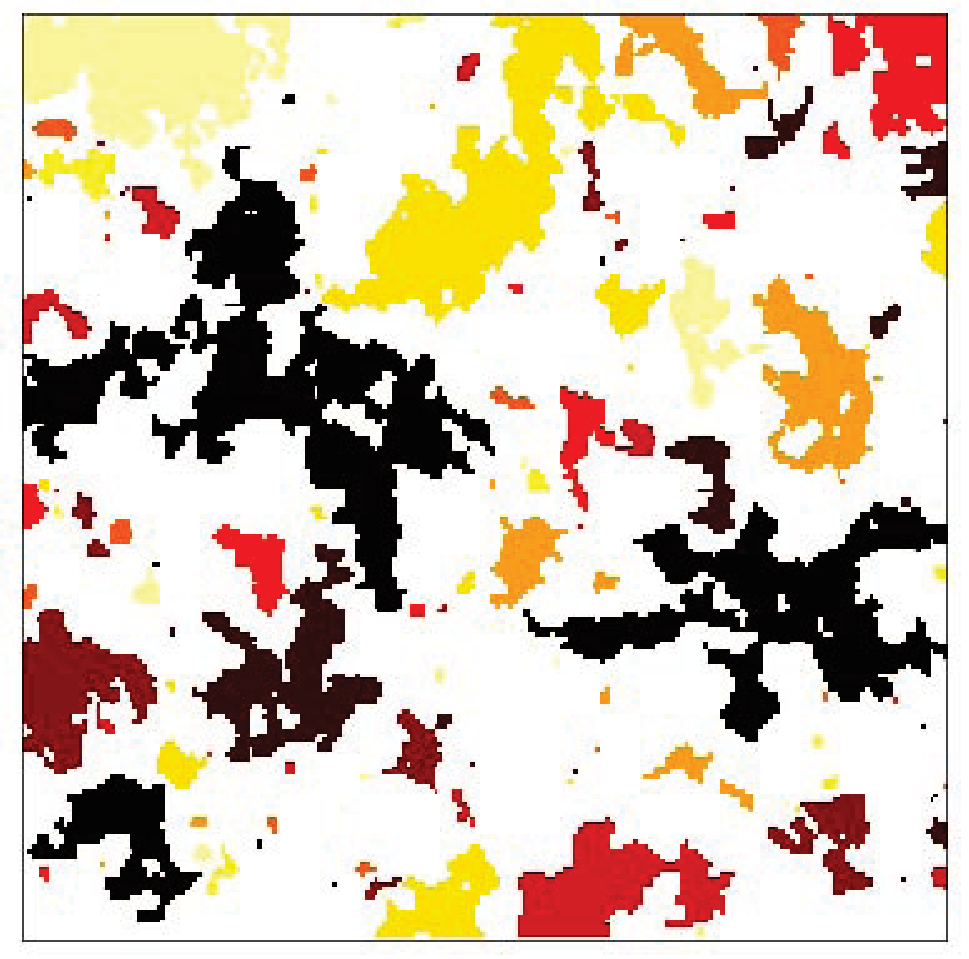}
		\caption{}
		\label{fig:mu01}
	\end{subfigure}
	\begin{subfigure}{0.3\textwidth}\includegraphics[width=\textwidth]{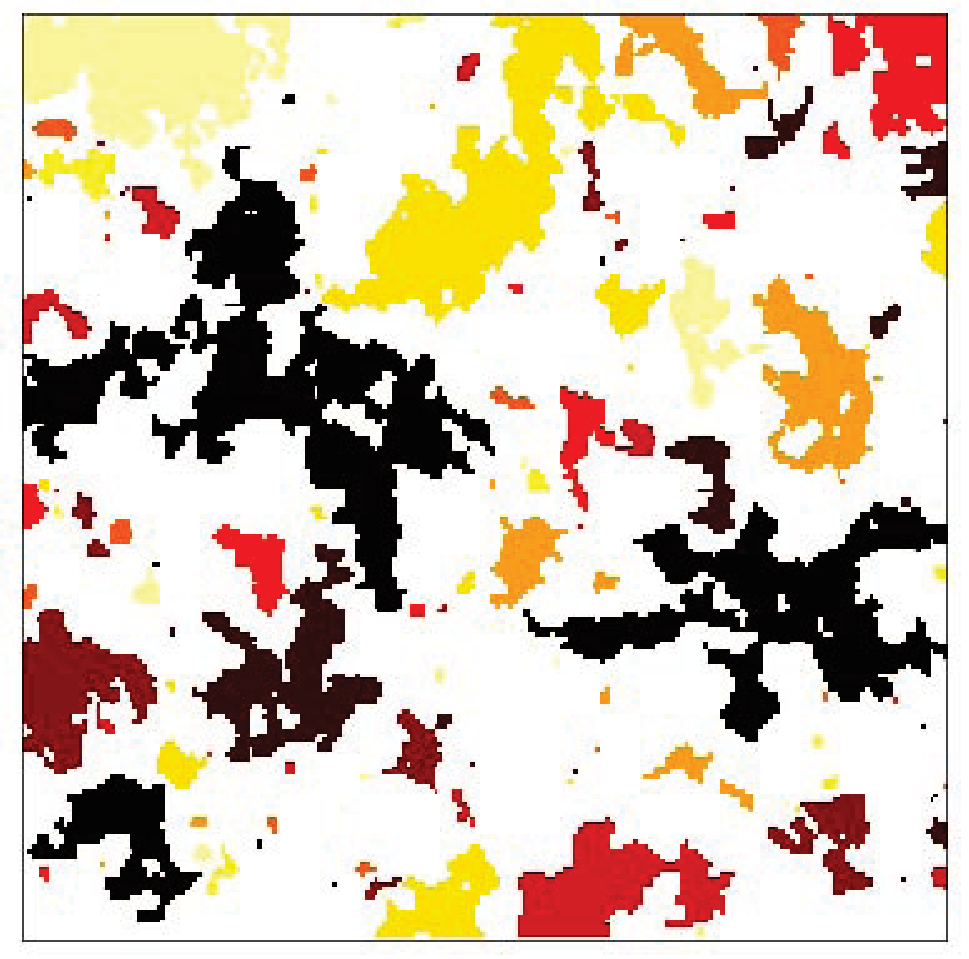}
		\caption{}
		\label{fig:mu05}
	\end{subfigure}
	\begin{subfigure}{0.3\textwidth}\includegraphics[width=\textwidth]{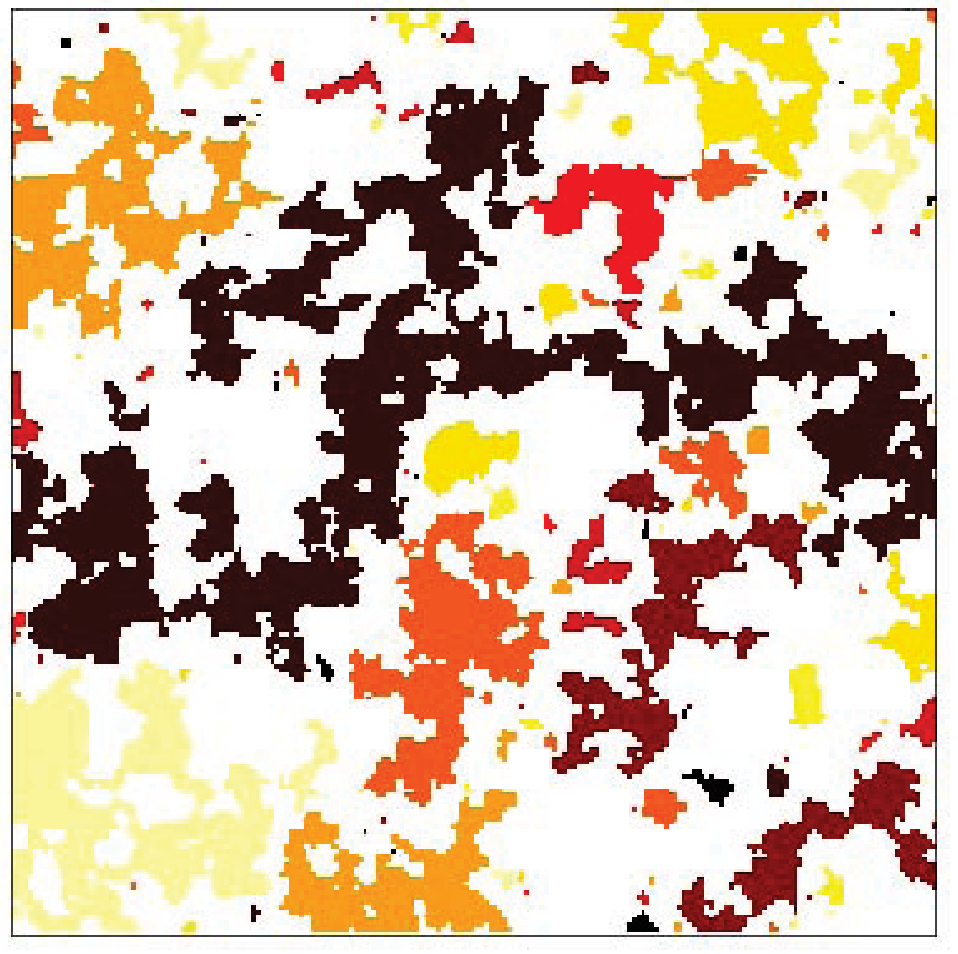}
		\caption{}
		\label{fig:mu1}
	\end{subfigure}
	\caption{(Color Online) The charge pattern of the systems with chemical potentials: (a) $\mu=0.001$ (b) $\mu=0.005$ (c) $\mu=0.01$ (d) $\mu=0.05$ (e) $\mu=0.1$ (f) $\mu=0.2$. The different colors show the connected components.}
	\label{fig:samples}
\end{figure*}

\begin{figure*}
	\begin{subfigure}{0.45\textwidth}\includegraphics[width=\textwidth]{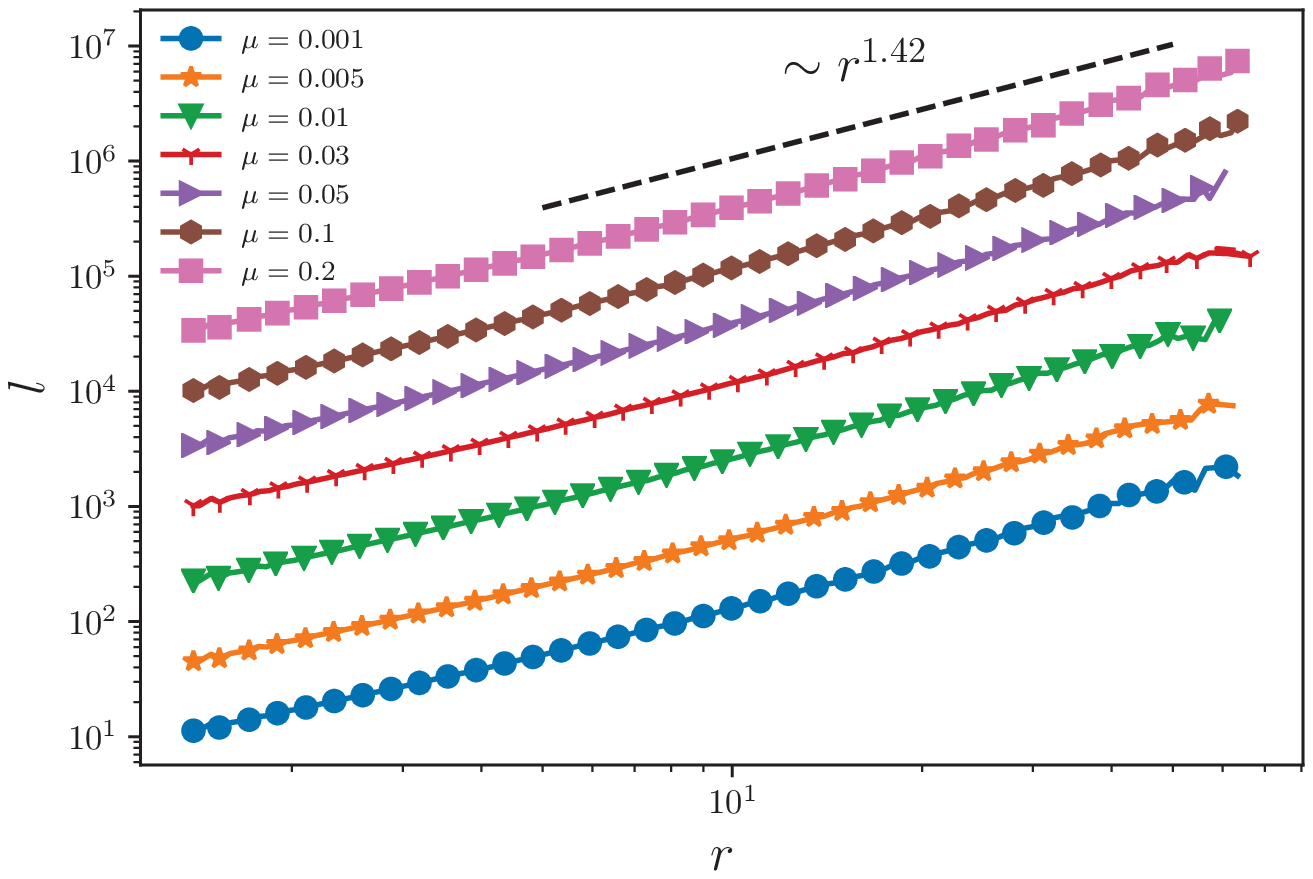}
		\caption{}
		\label{fig:l_r}
	\end{subfigure}
	\begin{subfigure}{0.45\textwidth}\includegraphics[width=\textwidth]{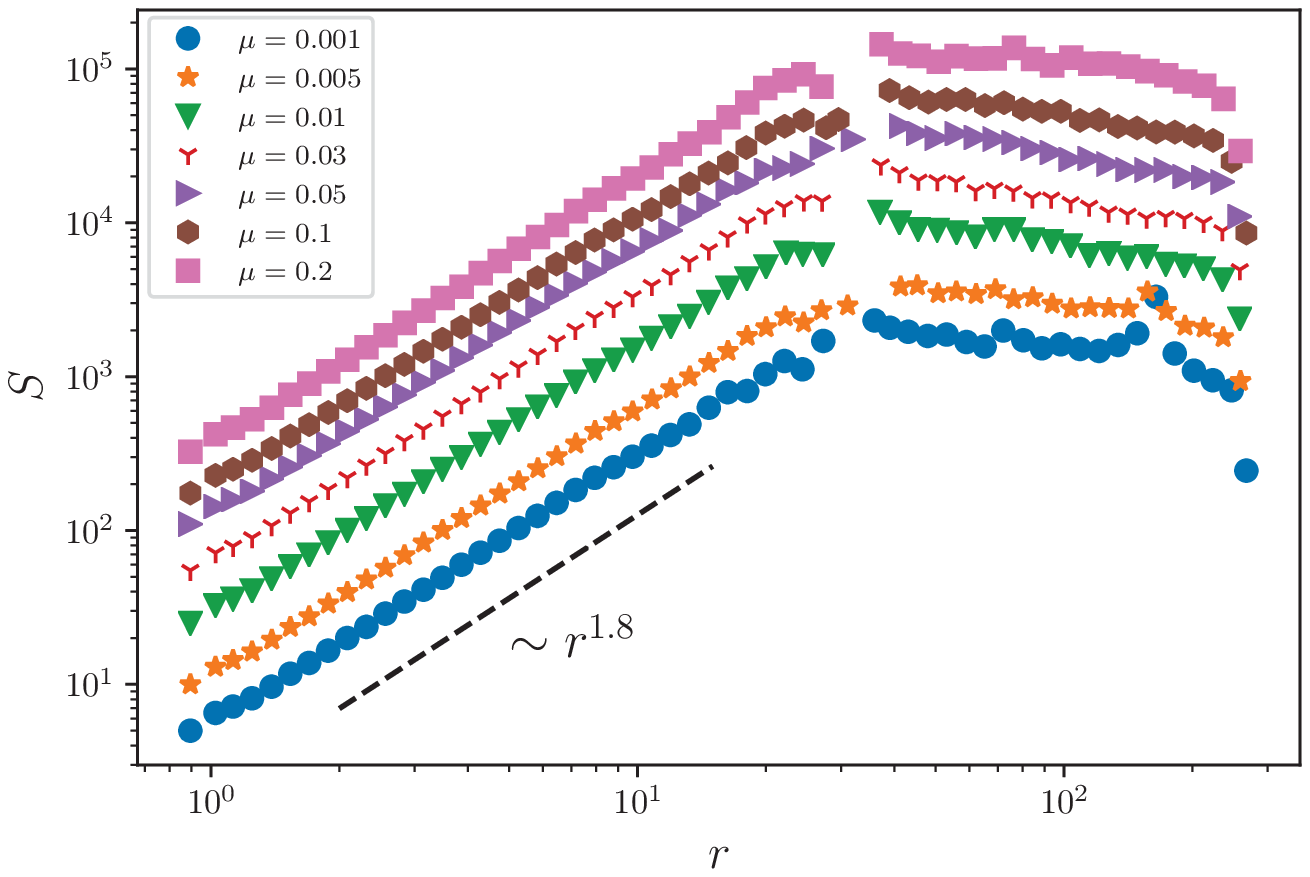}
		\caption{}
		\label{fig:S_R}
	\end{subfigure}
	\begin{subfigure}{0.45\textwidth}\includegraphics[width=\textwidth]{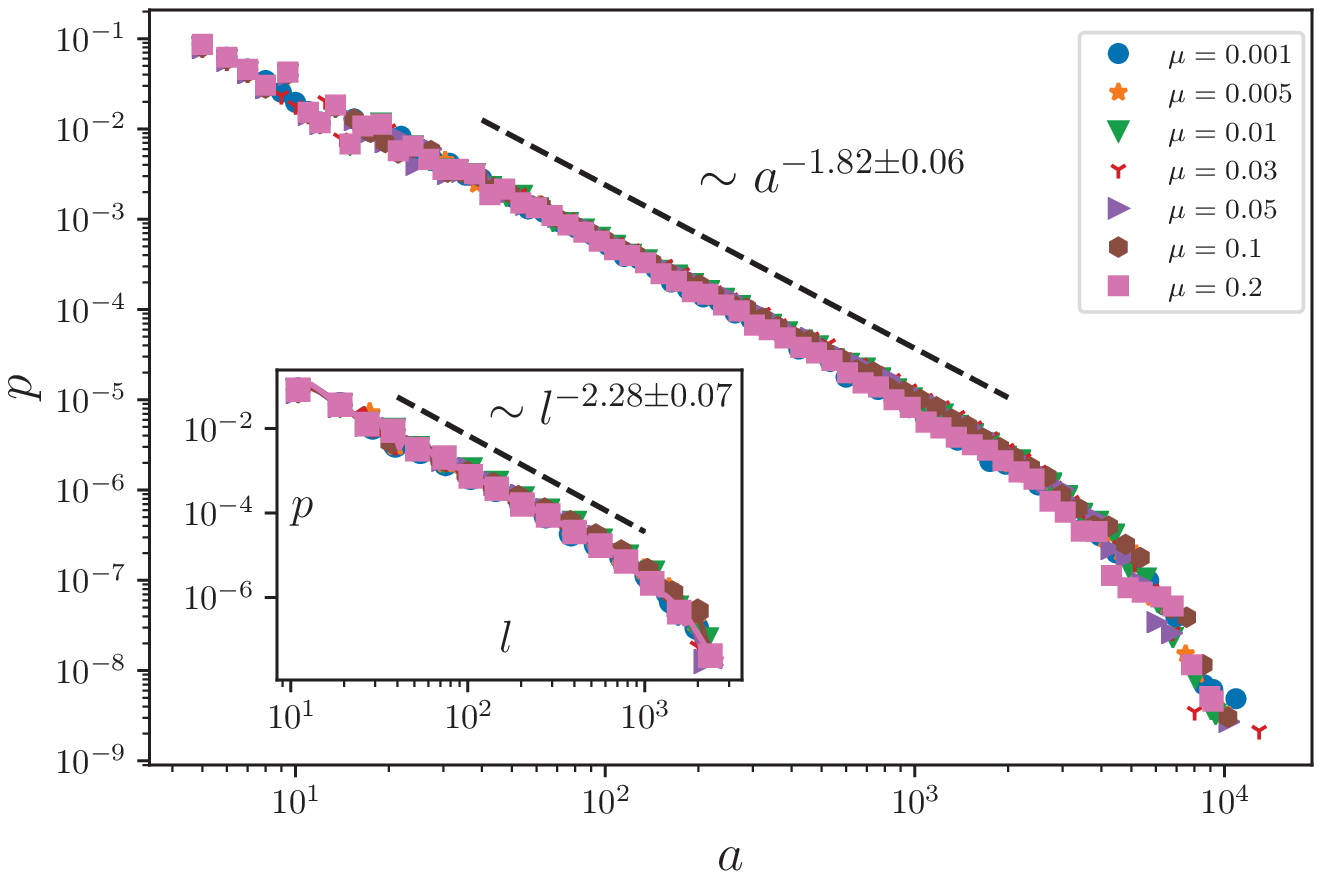}
		\caption{}
		\label{fig:pdf_l}
	\end{subfigure}
	\begin{subfigure}{0.45\textwidth}\includegraphics[width=\textwidth]{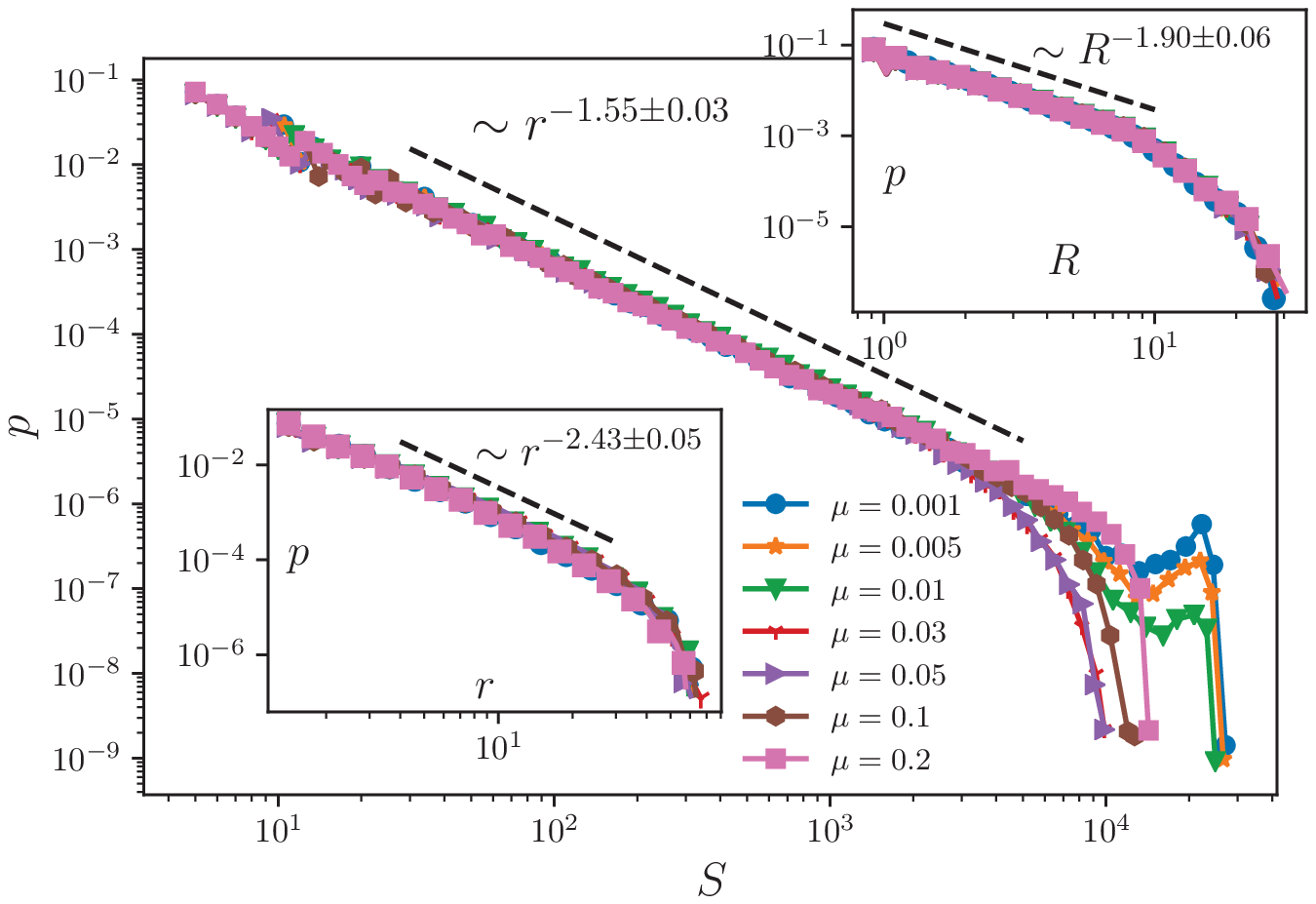}
		\caption{}
		\label{fig:pdf_S}
	\end{subfigure}
	\caption{(Color Online) The log-log plot of (a) $l$-$r$, (b) $S$-$r$ with the exponents $\gamma_{lr}=1.42\pm 0.02$ and $\gamma_{Sr}=1.8\pm 0.1$. (c) The distribution function of the area inside the cluster $a$ (inset: the distribution function of the loop length $l$). (d) The distribution function of the cluster mass $S$ (upper inset: the distribution function of the mass gyration radius $R$, lower inset: the distribution function of the loop gyration radius $r$). The corresponding exponents are: $\tau_{a}=1.82\pm 0.06$, $\tau_{l}=2.28\pm 0.07$, $\tau_{S}=1.55\pm 0.03$, $\tau_{R}=1.90\pm 0.06$ and $\tau_{r}=2.43\pm 0.05$.}
	\label{fig:geometrical}
\end{figure*}

The scale invariance of the critical systems dictate some scaling relations between the geometrical quantities of the system. The determination of the exponents of these global quantities, for example the various fractal dimensions of the system, helps in determining their universality class of the model in hand. All of the analysis presented in the previous section are in terms of local variable $n(\textbf{r})$. There is however a non-local point of view in such problems, i.e. the iso-height lines of the profile $n(\mathbf{r})$ at the level set $n(\mathbf{r}) = n_0$ which also show the scaling properties. When we cut the self-affine surface $n(x,y)$ some non-intersecting loops result which come in many shapes and sizes \cite{kondevprl,kondevpre}. We choose $10$ different $n_0$ between the maximum and the minimum densities from which a CLE is obtained. Some samples have been shown in Fig.~\ref{fig:samples} for $\mu=10^{-3},5\times 10^{-3},10^{-2},5\times 10^{-2},0.1$. The different colors in each figure show connected clusters each of which has its own gyration radius $r$, (exterior) loop length $l$, mass ($S$) and area inside ($a$) (this is the area inside the loop). For the self-affine systems, these geometrical objects are scale invariant and show various power-law behaviors, e.g., their size distribution is characterized by a few power law relations and scaling exponents. These quantities (in the thermodynamic limit) scale with each other in the form $y\sim x^{\gamma_{xy}}$ and the distribution functions of them behave like $p(x)\sim x^{-\tau_x}$ in which $x,y=l,r,S,a$. The scaling theory of CLEs of self-affine Gaussian fields was introduced in Ref.~\cite{kondevprl} and developed in Ref.~\cite{kondevpre,kondevother}. When a charge density pattern is obtained, we extract the contour lines by $10$ different cuts with the same spacing between maximum and minimum values. The Hoshen-Kopelman~\cite{hoshen1976percolation} algorithm has been employed for identifying the clusters in the lattice. It is notable that for each $L=200$ sample (for a given $\mu$) about $\sim 5\times 10^2$ loops were obtained. Since we have generated $4\times 10^3$ samples for each $\mu$, about $2\times 10^6$ loops have been generated for each $\mu$.\\
We have calculated the fractal dimensions $\gamma_{lr}$ and $\gamma_{Sr}$ as are seen in Figs.\ref{fig:l_r} and~\ref{fig:S_R}. Interestingly we have observed that these exponents are $\mu$-independent. The numerical values of these universal quantities are $\gamma_{lr}=1.42\pm 0.02$ and $\gamma_{Sr}=1.80\pm 0.05$. The latter changes behavior (more precisely becomes nearly constant) for the scales larger than $r_0\sim 30$ which is finite size effect. It is notable that for a space-filling cluster $\gamma_{Sr}^{\text{space filling}}=2$. The difference between the obtained $\gamma_{Sr}$ and $\gamma_{Sr}^{\text{space filling}}$ shows that there are some hallows (the regions with different densities) inside the clusters. The exponent $\gamma_{lr}$ is not significantly different from the case $\mu=0$~\cite{Najafi2017scale}. The exponents of the distribution function of the geometrical observables have also been presented in Figs.~\ref{fig:pdf_l} and~\ref{fig:pdf_S}. We see that $\tau_{a}=1.82\pm 0.06$, $\tau_{l}=2.28\pm 0.07$, $\tau_{S}=1.55\pm 0.03$, $\tau_{R}=1.90\pm 0.06$ and $\tau_{r}=2.43\pm 0.05$. These scaling behaviors for the geometrical quantities of the system which is surely not scale-invariant, are very interesting. For the critical systems, it is well-known that $\gamma_{xy}=\frac{\tau_y-1}{\tau_x-1}$. The fact that this hyper-scaling relation is not true for our system ($\gamma_{lr}^{\text{hyper-scaling}}\equiv\frac{\tau_r-1}{\tau_l-1}=1.12$ and $\gamma_{Sr}^{\text{hyper-scaling}}\equiv\frac{\tau_r-1}{\tau_S-1}=2.6$) is due to the fact that our system is not self-affine. It is notable that the obtained value for $\tau_l$ is compatible with the un-gated graphene ($\mu=0$). 

\section{Conclusion}
In this paper we have considered the graphene out of (and in the vicinity of) the Dirac point, i.e. for the finite chemical potentials. To this end we have employed the Thomas-Fermi-Dirac theory and have solved it numerically for finite $\mu$s. For the graphene system, out of the Dirac point, the impurity potential in not bare Coulomb potential, but instead is of the form of Eq.~\ref{Eq:potential}. We have mapped the problem to a rough surface system and have calculated the relevant functions which are of the scaling form for the rough surfaces. Importantly we have calculated multi-point charge correlation functions ($D_n$), the roughness function ($W_n$) ($n=2,3$ and $4$), and also the Fourier power spectrum function ($S_q$). \\
Our main finding is that these functions is decomposable to two parts and the function of $\mu$ is factored out. $D_n$ and $W_n$ show double-logarithmic behaviors with $r$ and $L$, whereas $\ln S_q$ has a linear behavior of the third power of $\ln q$. The $\mu$-functions behave in most cases with second power of $\mu$.\\
We have also analyzed the geometrical functions. To this end, we have cut the samples from some equal-spacing planes from which non-intersecting stochastic curves and loops result. Each loop has its own length ($l$), gyration radius ($r$), cluster mass ($S$), area inside ($a$). Our observations support that, despite the fact that the system is not self-affine, these quantities show critical (power-law) behaviors, and the resulting critical exponents are $\mu$-independent. These exponents are completely compatible with the corresponding exponents for un-gated graphene ($\mu=0$), i.e. geometrical properties of the system are independent of $\mu$ and also the spatial scale $1/q_{\text{TF}}$. Therefore, although $\mu$ changes the mean density $\bar{n}$, the geometrical properties of the system which is determined with respect to the new reference $\bar{n}(\mu)$ are robust against $\mu$. It can be understood noting the fact that all functions of $r$ (spatial distance) and $\mu$ are decomposable to two pure function, leading to the fact that the spatial structure of the system does not change by $\mu$.



\end{document}